\documentclass[12pt]{article}
\usepackage[centertags]{amsmath}
\newcommand{\be}{\begin{equation}}
\newcommand{\ee}{\end{equation}}
\newcommand{\bea}{\begin{eqnarray}}
\newcommand{\nn}{\nonumber}
\newcommand{\eea}{\end{eqnarray}}
\newcommand{\ti}{\widetilde}
\usepackage[centertags]{amsmath}
\usepackage{amsfonts}
\usepackage[]{epsfig}
\begin{document}
\title{Explicit R-Symmetry Breaking and Metastable Vacua}
\author{\normalsize  D.~Marqu\'es\thanks{Associated with
CONICET}$\ $ and F.~A.~Schaposnik\thanks{Associated with CICBA}
\\
{\small\it  Departamento de F\'{\i}sica, Universidad Nacional de La
Plata, C.C. 67, 1900 La Plata, Argentina}}

\maketitle
\begin{abstract}
We consider O'Raifeartaigh-like models with explicit R-symmetry
breaking and analyze the vacuum landscape. Taking such models as
candidates for the hidden sector, we analyze the gauge mediation of
the supersymmetry breaking, focusing on the effects produced by
R-symmetry breaking. First, we construct families of non-R-symmetric
models containing only singlet chiral superfields, and determine the
conditions under which SUSY vacua, runaway directions and
(longlived) metastable vacua exist. We then extend the results to
the case in which some of the chiral fields are in the $5 \oplus
\bar 5$ representation of $SU(5)$. Gauging this symmetry, we compute
soft masses for gauginos and sfermions,  and analyze several issues
such as doublet/triplet splitting, unification of coupling constants
and CP violation phases.
\end{abstract}

\section{Introduction}

The fact that supersymmetry (SUSY) breaking is intimately tied  to
the existence of a global $U(1)$ R-symmetry  was first stressed in
the work of Nelson and Seiberg \cite{Nelson:1993nf}, where it was
shown that in order to have SUSY breaking in generic models there
must be an R-symmetry, and the spontaneous breaking of this latter
symmetry is a sufficient condition for the existence of non
supersymmetric vacua. As it is well known, if supersymmetry is
realized in nature, it must be broken. Concerning R-symmetry,
generation of gaugino masses requires that it should be broken,
either explicitly or spontaneously.

It was recently shown in ref.~\cite{ISS1} that metastable
supersymmetry breaking is generic in supersymmetric field theory and
highly simplifies model-building. Moreover, there is a growing
consensus on the fact that such metastable supersymmetry breaking
takes place in a hidden sector of the full theory, and its effects
are communicated to the visible sector (MSSM) through gauge
interactions with messenger fields (this is the gauge mediation
scenario \cite{Dine:1981za}-\cite{Dine:1995ag}, see for example
\cite{Giudice:1998bp} for a review).

It should be stressed that metastability is closely related to
R-symmetry breaking. Concerning explicit R-symmetry breaking, stable
SUSY breaking vacua can become metastable in generic models since in
that case SUSY vacua can appear
\cite{Nelson:1993nf}. Metastable vacua also exist
when R-symmetry is spontaneously broken, for example by assigning
generic R-charges to fields \cite{Shih:2007av} or by coupling the
model to some broken gauge symmetry \cite{Intriligator:2007py}, \cite{Dine:2006xt}. There is yet another possibility in the field of R-breaking which is
that pseudo-moduli are only sensitive to two-loops in perturbation
theory \cite{Giveon:2008wp}.

As discussed in \cite{ADJK}, the way in which R-symmetry  is broken
(explicitly or spontaneously)  leaves a clear imprint on the
phenomenology of the minimal supersymmetric standard model (MSSM)
and it is  then of interest to study broad classes of such models so
as to compare the resulting patterns.

In the present paper we shall follow the gauge mediation route,
choosing for the hidden sector (non-generic) O'Raifeartaigh-type
models \cite{O'Raifeartaigh:1975pr}. R-symmetry will be broken
explicitly and in this sense our study can be seen as complementary
to that of ref.~\cite{Cheung:2007es}, where the R-breaking mechanism
is spontaneous. We start by constructing families of non-R-symmetric
models containing only singlet chiral superfields to describe the
hidden sector. A detailed analysis of the vacuum landscape will
allow us to determine the conditions under which SUSY vacua, runaway
directions and longlived metastable vacua exist.

In order to promote these models to more realistic candidates for
the SUSY breaking sector, we also consider the case in which some of
the chiral fields are in the $5 \oplus \bar 5$ representation of
$SU(5)$, leaving the field which triggers SUSY breaking as a singlet
spurion. By gauging the $SU(5)$ flavor symmetry, we allow some
fields to interact through gauge loops with the MSSM fields. We
compute soft masses and analyze issues such as doublet/triplet
splitting, unification of coupling constants and CP violation
phases.

The paper is organized as follows. In section 2 we discuss three
families of non-R-symmetric O'Raifear\-taigh-like models, having
different messenger mass matrices and   all chiral superfields
(messengers and spurion) taken as singlets. Then, in section 3 we
extend the analysis by considering the case in which messengers are
taken in the $ \bar5 \oplus 5$ representation of $SU(5)$. Section 4
addresses to the analysis of gauge mediation. Finally, in section 5
we summarize and discuss our results.

\section{O'Raifeartaigh-like models with explicit R-symmetry breaking}

The O'Raifeartaigh model  \cite{O'Raifeartaigh:1975pr}, a paradigm
of SUSY breaking, is a theory  with  three chiral superfields $X$,
$\phi_1$, $\phi_2$ transforming under a global $U(1)_R$
\be
\Phi(\theta) \to e^{i R(\Phi) \alpha} \Phi(e^{-i\alpha} \theta) \ ,
\ee
with charges $R(X) = 2,\ R(\phi_1) = 0,\ R(\phi_2) = 2$, a canonical K\"ahler
potential  and a superpotential of the form
\be W_{O'R} = \frac{\lambda}{2} X \phi_1^2 + m \phi_1 \phi_2 + f X\
. \label{ORaifeartaigh}\ee
Here $\lambda, m, f$ are complex. The model is renormalizable,
R-symmetric and not generic (not all the terms consistent with
R-symmetry are present, for example  $X \phi_1,\ \phi_2 \phi_1^2, \
\phi_2$ are absent). The field $X$, usually called a spurion,
 acquires a non-vanishing F-component, triggering SUSY
 breaking\footnote{as usual, we denote the superfield and its lowest
component with the same letter.}
\be
 X   = X + \theta^2 F\ . \label{VEVX1}
 \ee
{  When $\lambda f < m^2$ there is a phase in which } supersymmetry is spontaneously broken at
\be \phi_1 = \phi_2 = 0 \, , \ \ \ \ F = f \, , \ \ \ \ \forall\ X\ , \ee
and there are neither supersymmetric vacua nor runaway directions.
At the  tree level there is a one-dimensional  moduli space of
degenerate non-supersymmetric vacua parameterized by $X$. This
result is general, non-SUSY tree-level vacua is always degenerate in
Wess-Zumino models \cite{Ray:2006wk}-\cite{Sun:2008nh}. The
degeneracy is then lifted \`a la Coleman-Weinberg
\cite{Coleman:1973jx} when quantum corrections are taken into
account. At one loop, the vacuum expectation value (VEV) of the $X$
field vanishes (as it happens for $\phi_1$ and $\phi_2$) and then
the R-symmetry is not spontaneously broken. {  In addition there is another phase when $\lambda f > m^2$ containing two disjoint pseudo-moduli spaces, also lifted in such a way that the R-symmetry remains unbroken.}
We shall not analyze this last phase of the O'Raifeartaigh model.

In order to test different possibilities of SUSY breaking, this
model has been generalized in many ways, by adding fields and/or
considering generic superpotentials. For example, in
\cite{Intriligator:2007py} the following superpotential with $r$
fields $X_i$ and $s$ fields $\phi_j$ with R-charges $2$ and $0$
respectively has been proposed
\be W = \sum_{i=1}^r X_i\ g_i(\phi_j)\ , \ \ \ \ \ j = 1 ,\dots, s\ . \ee
Here the functions $g_i$ are generic and supersymmetry is broken
with an $(r-s)$-dimensional moduli space of non-supersymmetric vacua
parameterized by $X_i$. As in the previous case, this
 degeneracy is lifted at the quantum level in such a way that R-symmetry remains
unbroken. As shown in \cite{Intriligator:2007py}, coupling the $X$
field to a broken gauge symmetry leads to a vacuum with $
X_i \neq 0$, thus breaking R symmetry spontaneously. The
possibility of explicit (and small) R-symmetry breaking was
considered in that paper \cite{Intriligator:2007py} and in
\cite{Abe:2007ax}.

In \cite{Shih:2007av}, a particular generalization to the O'Raifeartaigh
superpotential (\ref{ORaifeartaigh}) was presented
\be W = f X + \frac{1}{2}(M_{ij} + X N_{ij})\phi_i\phi_j\ , \;\;\;\; i,j=1,2,\ldots, N_\phi \ .
\label{Superpotencial}
\ee
Here  $f \neq 0$ is a complex parameter, $M$ and $N$
symmetric complex matrices and $\det M \neq 0$. { For future reference, we define $N_R$ as the number of distinct R-charges carried by the $\phi$-fields. } R-symmetry is
guaranteed by requiring $R(X)=2$ and constraining $M$ and $N$ in
such a way that
\bea &&{\rm if~} M_{ij} \neq 0\ {\rm ~one~has~} R(\phi_i) +
R(\phi_j) = 2\ ,\nn\\&& {\rm if~} N_{ij} \neq 0\ {\rm ~one~has~}
R(\phi_i) + R(\phi_j) = 0\ . \label{RSymmCond}\eea
These selection rules imply that the following identity holds (see
Appendix)
\be \det(M + X N) = \det(M)\ .\label{DetNull} \ee
The SUSY vacua conditions are
\bea f + \frac{1}{2}\ N_{ij}\
\phi_i \phi_j  &=& 0 \ , \label{NoSUSYCond1}\\
(M + X N)_{ij}\ \phi_j &=& 0 \ .\label{NoSUSYCond3} \eea
As $\det M\neq 0$, equation (\ref{DetNull}) implies that
(\ref{NoSUSYCond3}) is only satisfied when $\phi_j = 0$  and then
SUSY is broken because it is not possible to satisfy
(\ref{NoSUSYCond1}) when the parameters are not fine-tuned.
R-symmetry is responsible of the selection rules (\ref{RSymmCond}),
which imply the identity (\ref{DetNull}), in which the r.h.s. is
non-vanishing by definition. Then, R-symmetry is a sufficient
condition for SUSY breaking, but it is not necessary since one
considers a non-generic model. There is always a SUSY-breaking vacua
at $\phi_i = 0$, with a moduli space parameterized by $X$. When only
R-charge $0$ and $2$ fields are present, one can show that there is
a minimum of the Coleman-Weinberg effective potential at $X = 0$,
and then R-symmetry remains unbroken at the quantum level (it was
shown in \cite{Ray:2007wq} that the comparison between the number of
R-charge 0 and R-charge 2 is essential to the symmetry breaking
properties of the model). Unbroken R-symmetry is no longer true when
the possibility for generic R-charge assignments (satisfying
conditions (\ref{RSymmCond})) is allowed. Models with superpotential
(\ref{Superpotencial}) were generalized to the case of many
pseudo-moduli fields in \cite{Ferretti:2007ec}, where it was also
shown that they generically have runaway directions. The case of
non-canonical K\"ahler potentials was analyzed in
\cite{Aldrovandi:2008sc}.

We will consider here a different extension of these models allowing
the possibility of explicit R-symmetry breaking. As it is well-known
\cite{Nelson:1993nf},  in generic models when R-symmetry is
explicitly broken SUSY vacua always exist and those breaking SUSY
could be in principle close to them. Then, since we are interested
in longlived supersymmetry breaking vacua, we shall consider
non-generic models with a controllable life-time. The model is
defined by a canonical K\"ahler potential and the (non-generic)
superpotential
\be W = f X + \frac{1}{2}(M + B + X (L + A))_{ij}\phi_i\phi_j \, ,
 \;\;\;\;\;\;\;\;\;\;\;\; i,j=1,2,\ldots,N_\phi \ . \label{SuperpotRBreak} \ee
Here  $f \neq 0$ is a complex para\-meter and $M$,
$B$, $L$ and $A$ are symmetric complex matrices satisfying
\bea M_{ij} \neq 0 &\Rightarrow& R(\phi_i) +
R(\phi_j) = 2\ , \nonumber \\
 L_{ij}\neq 0 &\Rightarrow& R(\phi_i) +
R(\phi_j) = 2\ , \nonumber \\
B_{ij} \neq 0  &\Rightarrow& R(\phi_i) +
R(\phi_j) < 2\ ,\nonumber\\ A_{ij} \neq 0  &\Rightarrow& R(\phi_i) +
R(\phi_j) < 2\ . \label{SelectionRules}\eea
As before, we define $N_R$ as the number of distinct R-charges carried by the $\phi$-fields. Here we consider the same charge assignment as in (\ref{Superpotencial}) and then,
as $B$, $L$ and $A$  in  superpotential  (\ref{SuperpotRBreak}) satisfy selection rules
(\ref{SelectionRules}) and not those required
by   R-symmetry (\ref{RSymmCond}), the  models defined
by (\ref{SuperpotRBreak})-(\ref{SelectionRules}) in general
explicitly break R-symmetry. Only when $ B = L = 0$, $A_{ij}\ne 0$ requires $R(\phi_i) + R(\phi_j) = 0$, and 
$\det M \neq 0$, the superpotential  $W$ coincides with
(\ref{Superpotencial}) and then the model is R-symmetric. In general
a reassignment of R-charges might be necessary in order to check
whether other choice of matrices renders the model R-symmetric (An
example of this is discussed in the Appendix).

The scalar potential resulting from (\ref{SuperpotRBreak}) is
\be V = | f + \frac{1}{2} (L + A)_{ij}\ \phi_i \phi_j |^2 + |(M
+ B + X (L + A))_{ij} \ \phi_j |^2 \ ,
\label{formula}\ee
and then the F-term conditions for SUSY vacua read
\bea f + \frac{1}{2} (L + A)_{ij}\ \phi_i \phi_j &=& 0\ ,
\label{not1}\\
(M + B + X (L + A))_{ij} \ \phi_j &=& 0\ .
\label{not2}\eea
There is a local (classical) extrema at $\phi_i = 0\ ,\ \forall\,
X$, in which $V = |f|^2$. At one loop, a Coleman-Weinberg potential
\cite{Coleman:1973jx} is generated on the pseudomoduli and the
minima of the resulting effective potential, if they exist, will be the
SUSY-breaking vacua of the theory. Notice also that if the effective potential
lifts the moduli in such a way that the true minima is at $X=0$, then the R
symmetry would be restored. Hence the present models
provide an appropriate context to analyze  R-symmetry
restoration.

In principle, those SUSY breaking minima can be made longlived in
the presence of SUSY vacua or runaway directions. In fact, when $X$
is such that $\det(M + B + X (L + A)) \neq 0$ there is no SUSY vacua
because (\ref{not2}) implies that $\phi = 0$, which makes
(\ref{not1}) unsolvable. In contrast, when $X$ is such that $\det(M
+ B + X (L + A)) = 0$, there is a non-zero $\phi$ solution to
eq.(\ref{not2}) (and hence a SUSY vacuum)
 which can
be taken far away for small entries of $L$ and $A$ in (\ref{not1}). The mass matrices at the extrema read
\bea {\cal M}_B^2 \!\!\!\!&=&\!\!\!\! \left(\begin{matrix}
W_{ik}^\dag W^{kj} & W_{ijk}^\dag W^k \\ W^{ijk} W_k^\dag & W^{ik}
W_{kj}^\dag
\end{matrix}\right) = (\hat M + X \hat L)^2 + f \hat L\ ,\nonumber\\
 {\cal M}_F^2 \!\!\!\!&=& \!\!\!\!\left(\begin{matrix} W_{ik}^\dag W^{kj} &
0 \\ 0 & W^{ik} W_{kj}^\dag
\end{matrix}\right) = (\hat M + X \hat L)^2\ ,\label{Masas}\eea
where
\be \hat M = \left(\begin{matrix} 0 & (M + B)^\dag \\ M + B
& 0
\end{matrix}\right)\ , \ \ \ \ \hat L = \left(\begin{matrix} 0 & (L + A)^\dag \\ L + A & 0
\end{matrix}\right) \ .\label{masas2}\ee
The stability of the moduli space is guaranteed as long as these
matrices have no tachyonic eigenvalues (the stability of the vacuum strongly depends on the R-symmetry breaking \cite{Abe:2007ax}). As wee will see,
 in general one can take small entries for $L$ and $A$ without
 destabilizing the non-supersymmetric vacuum, and thus one can in
 principle make it longlived. In addition to the non-SUSY vacua at the origin of field space,
  there can be other
non-supersymmetric minima elsewhere.

Let us label fields so that $i<j \Rightarrow R(\phi_i)\leq
R(\phi_j)$. There will be $(n_1, \dots, n_{N_R})$ fields with charge
assignments $(r_1, \dots, r_{N_R})$ ordered in increasing order. We
group them in $N_R$ vectors $\phi^{(r_i)}_{a_i}$, each one having
  $n_i$ components ($a_i = 1,\dots,n_i$). Matrices $M$,
$B$, $L$ and $A$ can then be arranged in  blocks of $n_i$ rows and
$n_j$ columns labeled by R-charges which will be denoted as $(M + B
+ X (L + A))^{(r_i, r_j)}_{a_i, a_j}$ Sometimes we will omit indices
$a_k$ to simplify notation. The models we shall consider have
$\det(M + X L) \neq 0$ for some values of $X$. Then, as we explain
in the Appendix, in the basis in which the fields are ordered by
increasing R-charge, $M + X L$ in anti-diagonal by blocks of
non-zero determinant (except for some particular values of $X$), and the fields must come in pairs with
R-charges $r_i + r_{N_R - i + 1} = 2$. In this basis $\cal M$ has
the form

\be {\cal M}(X) = \left(\begin{matrix} (B + X A)^{(r_1, r_1)} & &
\dots & & (B + X A)^{(r_1, r_{N_R-1})} & (M + X L)^{(r_1, r_{N_R})} \\  &  & & & & 0 \\ \vdots & & \dots & & & \\ & & & &\dots & \vdots\\
(B + X A)^{(r_{N_R-1}, r_1)} & & & & & \\
(M + X L)^{(r_{N_R}, r_1)} &  & 0& & \dots& 0\end{matrix}\right)
\label{AntiDiagonal1}\ . \ee

\subsection{SUSY vacua, runaway directions and stability}

We will discuss here 3 families of models, each one with the same
vacuum structure. This classification is inspired on that made in
\cite{Cheung:2007es}, and the runaway analysis is similar to that in \cite{Ferretti:2007ec}. Before presenting the detailed analysis, let
us define each class and advance the main properties of their
corresponding vacua, which in the three cases can be made
parametrically longlived.

\begin{itemize}
\item \underline{Type I models}:  $L = 0$ and $\det M \neq 0$.   SUSY is spontaneously broken,
 but there exists a runaway direction, $V \to 0$ for $|X|\to\infty$ when $N_R > 2$,
 and  the $\phi$ fields behave asymptotically as
    \be |\phi^{(r_k)}_{a_k}| \sim |X|^{\frac{2k - N_R}{2}} \ , \ \ \ \ \forall\ k = 1, \dots, N_R\ , \ \ \forall\ a_k = 1, \dots, n_k\ .
    \ee
The  non-supersymmetric configuration $\phi = 0$ is a stable
minimum in some region $|X| < X_{max}$ when $f \ll eigenvalues(\hat M ^2)$.

\item \underline{Type II models}:  $M = 0$ and $\det L \neq 0$.  SUSY is spontaneously broken,
 but there exists a runaway direction at $X\to 0$
 when $N_R > 1$, in which the fields behave asymptotically as
    \be |\phi^{(r_k)}_{a_k}| \sim |X|^{\frac{N_R-2k + 1}{2}} \ , \ \ \ \ \forall\ k = 1, \dots, N_R\ , \ \ \forall\ a_k = 1, \dots, n_k\ .
    \ee
The  non-supersymmetric configuration $\phi = 0$ is a stable
minimum in some region $X_{min} <  |X|$.

\item \underline{Type III models}:  $\det M = 0, \det L = 0,
 \det(M + L) \neq 0$ and $M^{(r_i, r_{N_R-i+1})} = 0 \Leftrightarrow L^{(r_i, r_{N_R-i+1})}
\neq 0$. Calling $N_M(k)$ the number of non-zero blocks $M^{(r_i,
r_{N_R-i+1})}$ of the matrix $M$
 ($i = 1, \dots, k$), and $N_L(k) = k - N_M(k)$, then
\begin{itemize}
\item If $N_M(1) = 0$ the  models
have non-generic SUSY vacua at  $X = 0$  (for fine tuned values of the
parameters).
\item If $N_M(1) = 1$ there is always a SUSY vacuum at $X=0$
for finite values of the fields (no fine tuning).

\item If $N_M(N_R) > 2$ there is always a (Type I)
runaway direction parameterized by $|X|\to \infty$, in which the
fields behave asymptotically as \be |\phi^{(r_k)}_{a_k}| \sim
|X|^{\frac{2 N_M(k) - N_M(N_R)}{2}}  \ , \ \ \ \ \forall\ k = 1,
\dots, N_R\ , \ \ \forall\ a_k = 1, \dots, n_k\ .\ee

\item  If $N_L(N_R) > 1$
there is always a (Type II) runaway direction parameterized by $X\to
0$, in which the fields behave asymptotically as \be
|\phi^{(r_k)}_{a_k}| \sim |X|^{-\frac{2 N_L(k) - N_L(N_R) - 1}{2}} \
, \ \ \ \ \forall\ k = 1, \dots, N_R\ , \ \ \forall\ a_k = 1, \dots,
n_k\ . \ee
\end{itemize}

The  non-supersymmetric configuration $\phi = 0$ is a stable
minimum in some region $X_{min} <  |X| < X_{max}$ when $f \ll eigenvalues(\hat M ^2)$.
\end{itemize}

The particular form of the matrices   (\ref{SelectionRules}) is
imposed so that there is no SUSY vacua population. If one adds an
R-symmetry breaking term not respecting (\ref{SelectionRules}),
there would be a supersymmetric vacuum in a finite region of field
space in which $X \neq 0$, as it happens in the explicit R-breaking
example of \cite{Intriligator:2007py}.

Here  $A \sim B$  indicates  that $A$ and $B$ are proportional in a
given limit to be specified. Bounds $X_{min}$ and $X_{max}$ depend on the specific parameters of the model.
We will now prove these results (see the Appendix for specific
examples).

~

\noindent{\bf Type I models}

Type I models correspond to the case $L = 0$ and $\det M \neq 0$. In
these type of models supersymmetry is spontaneously broken. In fact,
the conditions for SUSY breaking read
\bea f + \frac{1}{2} A_{ij}\ \phi_i \phi_j &=& 0\ , \label{TypeIec1}\\
(M + B + X A)_{ij} \ \phi_j &=& 0\ , \label{TypeIec2}\eea
and we have proven in the Appendix that $\det(M + B + X A) = \det M
\ne 0$. Then, the only solution to (\ref{TypeIec2}) is $\phi_i = 0$,
which is inconsistent with (\ref{TypeIec1}). Notice in (\ref{Masas}) that as
$\det \hat M \neq 0$ and $\det \hat L = 0$, when $f \ll eigenvalues(\hat M ^2)$ all eigenvalues
of $m_B^2$ are real and positive in a neighborhood of $X = 0$ and this extrema is
a stable minimum.

In order to discover possible runaway directions, we first
demonstrate that the subset of equations
\bea
f + \frac{1}{2} A_{ij}\ \phi_i\phi_j &=& 0\ ,\label{firstRA}\\
(M + B + X A)_{ij}\ \phi_j &=& 0 \ , \ \ \ \ \ \ R(\phi_i) < r_{N_R}\ ,
\label{secondRA} \eea
can always be solved. To see this, let us rewrite the equations
(\ref{secondRA}) making explicit the R-charge of each field using
the notation explained above
\be \sum_{i= 1}^{k} (B + X A)^{(r_{N_R-k}, r_i)} \phi^{(r_i)} +
M^{(r_{N_R-k}, r_{k+1})} \phi^{(r_{k+1})} = 0 \ , \ \ \ \ \ \ \ \ k
= 1, \dots, N_R-1\ . \label{Solvable}\ee
For a given configuration $\phi^{(r_1)}$, these are $N_R-1$
(vectorial) equations for $N_R-1$ variables for every given $X$, so
they can generically be satisfied (the fact that $\det (M) \neq 0$,
implies that $\det (M^{(r_i, r_{N_R-i + 1})})\neq 0$). Notice that
given a configuration, a scaling of the fields $\phi_i \to \alpha
\phi_i$ is a solution to (\ref{Solvable}) as well. Then, we can
always find an $\alpha$ such that (\ref{firstRA}) is satisfied. This
shows that (\ref{Solvable}) together with (\ref{firstRA}) is the
largest subset of equations that can be satisfied if we look for
non-vanishing configurations.

The remaining equations
\be M^{(r_{N_R}, r_1)} \phi^{(r_1)} = 0\ , \label{UnSolvable}\ee
can only be solved for $\phi^{(r_1)} = 0$. Then,
(\ref{firstRA})-(\ref{secondRA}) are inconsistent with
(\ref{UnSolvable}) for finite and non-vanishing values of the
fields.

However, we can consider the limit $|X| \to \infty$ in such a way
that (\ref{Solvable}) are satisfied for non vanishing values of the
fields when $\phi^{(r_1)} \to 0$. Then, the only condition that
remains to be verified is (\ref{firstRA}). Notice that when $|X| \to
\infty$, generically we have $|\phi^{(r_k)}_{a_k}| \sim |X|^{k-1}
|\phi^{(r_1)}_{a_1}|\ , \ \forall\ k,\ a_k,\ a_1$. Then, in that
limit we can write in equation (\ref{firstRA})
\be \left|\sum_{i = 1}^{N_R-1} \sum_{j = 1}^{N_R - i}
(\phi^T)^{(r_i)} A^{(r_i, r_j)} \phi^{(r_j)}\right| \sim
\left|\sum_{i = 1}^{N_R-1} \sum_{j = 1}^{N_R - i} \|A^{(r_i, r_j)}\|
X^{i - 1} X^{j - 1} \phi^{(r_1)}_1\phi^{(r_1)}_1 \right|
 \ .\ee
The leading term when $|X| \to \infty$ is $X^{N_R-2}
(\phi^{(r_1)}_1)^2$, and if we make this term finite, all other
terms vanish in the expansion (avoiding possible divergent terms)
making equation (\ref{firstRA}) solvable . Then, we will have a
runaway direction if
\be |\phi^{(r_1)}_{a_1}| \sim |X|^{-(\frac{N_R-2}{2})}\ , \ \ \ \
a_1 = 1, \dots\ , n_1\ ,\ee
so we need $N_R > 2$ in order for $\phi^{(r_1)} \to 0$. The other
fields behave as $|\phi^{(r_k)}_{a_k}| \sim |X|^{\frac{2k -
N_R}{2}}\ , \ k = 1, \dots, N_R\ , \ a_k = 1, \dots, n_k$. Note that
asymptotically $|\phi^{(> r_{N_R/2})}| \geq |X| \to \infty$ so this
vacua is a runaway also in the $\phi-$field space.

~

\noindent{\bf Type II model}

Type II models have $M = 0$, $\det L \neq 0$. In these type of
models supersymmetry is spontaneously broken. In fact, the
conditions for SUSY breaking read
\bea f + \frac{1}{2} (L + A)_{ij}\ \phi_i \phi_j &=& 0\ ,\label{TypeIIec1}\\
(B + X (L + A))_{ij} \ \phi_j &=& 0\ . \label{TypeIIec2}\eea
We have proven in the Appendix that $\det(B + X (L + A)) = \det(X L)
= X^{N_\phi} \det L$, which is non-zero if $X \neq 0$. If this is
the case, the only solution to (\ref{TypeIIec2}) is $\phi_i = 0$,
which is inconsistent with (\ref{TypeIIec1}). As
$\det \hat L \neq 0$, this extrema is stable at sufficiently
large $|X|$.

Before turning to the study of the runaway behavior, let us briefly
show that generically there is no SUSY vacua for finite values of
fields when $X = 0$. Let us rewrite equations
(\ref{TypeIIec1})-(\ref{TypeIIec2}) as
\bea f + \frac{1}{2} (\phi^T)^{(r_i)}\ (L+A)^{(r_i, r_j)}\ \phi^{(r_j)} &=& 0\ , \label{ALequation}\\
B^{(r_i, r_j)} \ \phi^{(r_j)} &=& 0\ . \label{Bequation}\eea
Equation (\ref{Bequation}) is solved by $\phi^{(r_i)} = 0$ with $i =
1, \dots, N_R-1$, and any $\phi^{(r_{N_R})}$. Then, as $(L + A)^{(r_{N_R}, r_{N_R})} = 0$, equation (\ref{ALequation}) cannot be
satisfied and SUSY is broken.

Now we will demonstrate that the subset of equations
\bea
f + \frac{1}{2} (L + A)_{ij}\ \phi_i\phi_j &=& 0\ ,\label{firstRAII}\\
(B + X (L + A))_{ij}\ \phi_j &=& 0 \ , \ \ \ \ \ \ R(\phi_i) <
r_{N_R}\ , \label{secondRAII} \eea
can always be solved for non-vanishing $X$. Let us rewrite the
equations (\ref{secondRAII}) making explicit the R-charge of the
fields they involve with the notation we have introduced
\be \sum_{i= 1}^{k}(B + X A)^{(r_{N_R-k}, r_i)} \phi^{(r_i)} + X
L^{(r_{N_R-k}, r_{k+1})} \phi^{(r_{k+1})} = 0\ , \ \ \ \ \ \ \ \ k =
1, \dots, N_R - 1\ . \label{SolvableII}\ee
For a fixed $X \neq 0$, given a configuration $\phi^{(r_1)}$, these
are $N-1$ (vectorial) equations for $N-1$ variables, so they can
generically be satisfied (the fact that $\det L \neq 0$ implies
that $\det (L^{(r_i, r_{N_R-i + 1})})\neq 0$).

As in type I models, after some rescaling of the fields, the biggest
subset of equations that can be satisfied if we look for
non-vanishing configurations and $X \neq 0$ is (\ref{SolvableII})
together with (\ref{firstRAII}). The remaining equations
\be X L^{(r_{N_R}, r_1)} \phi^{(r_1)} = 0\ , \label{UnSolvableII}\ee
can only be solved when $X \phi^{(r_1)} = 0$. Then,
(\ref{firstRAII}),(\ref{SolvableII}) are inconsistent with
(\ref{UnSolvableII}) for finite and non-vanishing values of the
fields when $X \neq 0$. However, we still have the possibility of
taking $|X| \to \infty$ or $X \to 0$, in such a way that
(\ref{SolvableII}) is satisfied together with $X \phi^{(r_1)} \to
0$, in which case the only condition that would remain to be
verified is (\ref{firstRAII}). Let us analyze these two
possibilities.

$\bullet$ In the $|X| \to \infty$ case, in order for
(\ref{firstRAII}) to be satisfied we need $|\phi^{(r_k)}_{a_k}|\sim
|\phi^{(r_1)}_1|\ ,\ k = 1, \dots, N_R\ ,\ a_k = 1, \dots , n_k$.
Then if $\phi^{(r_1)} \to 0$ we violate the non-vanishing
requirement, and otherwise we violate the requirement $X
\phi^{(r_1)}\to 0$.

$\bullet$ In the $X \to 0$ limit, generically we have
$|\phi^{(r_k)}_{a_k}| \sim |X|^{1 - k} |\phi^{(r_1)}_1|\ ,\ k = 1,
\dots, N_R\ ,\ a_k = 1, \dots , n_k$. Then, in that limit, we can
write in (\ref{firstRAII})
\be \left| \sum_{i = 1}^{N_R} \sum_{j = 1}^{N_R - i + 1}
(\phi^T)^{(r_i)}(A + L)^{(r_i, r_j)} \phi^{(r_j)}\right| \sim \left|
\sum_{i = 1}^{N_R} \sum_{j = 1}^{N_R - i + 1} \|(A + L)^{(r_i,
r_j)}\| X^{1 - i} X^{1 - j} \phi^{(r_1)}_1\phi^{(r_1)}_1\right|\
.\ee

The leading term when $X \to 0$ is clearly $X^{1-N_R}
(\phi^{(r_1)})^2$, and if we make this term finite, all other terms
vanish in the expansion, and there will be no divergent terms. Then,
we will have SUSY vacua if
\be |\phi^{(r_1)}_{a_1}| \sim |X|^{\frac{N_R-1}{2}}\ , \ \ \ a_1 =
1, \dots, n_1\ ,\ee
because in this case $X \phi^{(r_1)} \to 0$ and (\ref{UnSolvableII})
is satisfied. But we know that there can not be SUSY vacua so this
limit must correspond to a runaway direction. In fact, the other
fields behave in this limit as $|\phi^{(r_k)}_{a_k}| \sim
|X|^{\frac{N_R - 2k + 1}{2}}$ $\forall\ k ,\  a_k$, and in
particular, $|\phi^{(r_{N_R})}_{a_{N_R}}| \sim |X|^{\frac{1 -
N_R}{2}} \to \infty\ , \ \forall\ a_{N_R}$, if $N_R
> 1$.

~

\noindent{\bf Type III}

Type III models have $\det M = 0$, $\det L = 0$ and $\det(M + L)
\neq 0$. These matrices are only non-zero in blocks of the type
$M^{(r_i, r_{N_R-i+1})}$, $L^{(r_i, r_{N_R-i+1})}$ with $i= 1,\dots,
N_R$, and we take them to satisfy
\be M^{(r_i, r_{N_R-i+1})} = 0 \Leftrightarrow L^{(r_i, r_{N_R-i+1})}
\neq 0 \ .\ee
Let us call $N_M(k)$ the number of non-zero blocks of the matrix $M$
from $i = 1, \dots, k$, and also define $N_L(k) = k - N_M(k)$.

The conditions for SUSY breaking read
\bea f + \frac{1}{2} (L + A)_{ij}\ \phi_i \phi_j &=& 0\ ,\label{TypeIIIec1}\\
(M + B + X(L + A))_{ij} \ \phi_j &=& 0\ . \label{TypeIIIec2}\eea
We have proven in the Appendix that $\det(M + B + X (L + A)) =
\det(M + X L) = X^n \det(M + L)$, which is non-zero when $X\neq 0$.
Here we have defined
\be
n = \sum_{i=1}^{N_R} n_i (N_L(i) - N_L(i-1)) \neq 0\ .
\ee
Then, the only possibility for SUSY vacua is taking $X = 0$. If $X
\neq 0$, the only solution to (\ref{TypeIIIec2}) is $\phi_i = 0$,
which is inconsistent with (\ref{TypeIIIec1}). Since $\det \hat M = 0$ and $\det \hat L = 0$, these models
share properties of type I and II models, and the non-SUSY minimum will typically be stable only in some range $X_{min} < |X| < X_{max}$.

~

Now we prove that when $X = 0$, $(i)$ if $N_M(1) = 0$ these models
have non-generic SUSY vacua (only for fine tuned values of the
parameters), and $(ii)$ if $N_M(1) = 1$ there is always a SUSY vacua
for finite values of the fields.

~

$(i)$ Consider the $(N_\phi - n_{N_R}) \times (N_\phi - n_{N_R})$
matrix formed by the $(r_i, r_j)$ blocks of $M + B$ with $i, j = 1,
\dots, N_R - 1$, and call it $H$. If $N_M(1) = 0$, then
$\phi^{(r_{N_R})}$ remains undetermined and the rest of the fields
$\phi^{(< r_{N_R})}$ have non-zero values only if $\det H = 0$. In
this case there can be SUSY vacua, but only for those fine-tuned
values of parameters. If $\det H \neq 0$ we have $\phi^{(< r_{N_R})}
= 0$, but in that case the equation (\ref{TypeIIIec1}) is not
solvable since $(L + A)^{(r_{N_R}, r_{N_R})} = 0$, so SUSY is
broken.

~

$(ii)$ If $N_M(1) = 1$, then $\phi^{(r_1)} = 0$. Moreover, there is
necessarily a $k$ in the range $1 < k \leq (N_R + 1)/2$ satisfying
$N_L(k) = 1$ and $N_L(k - 1) = 0$. For this $k$ we have $\phi^{(<
r_k)} = 0$, the fields $\phi^{(r_k)}$ are undetermined, and all
$\phi^{(> r_k)}$ depend on $\phi^{(r_k)}$. This implies that
equations (\ref{TypeIIIec2}) can always be solved. Regarding
equation (\ref{TypeIIIec1}), it depends on products of the form
$\phi^{(r_k)} \phi^{(\leq N_R-k+1)}$, but as $N_R > N_R - k + 1 \geq
(N_R + 1)/2$, there are non-vanishing terms depending on
$\phi^{(r_k)}$, and it can always be solved so there is always a
SUSY vacua.

~

In addition, these models have runaway behavior which we have
classified in two types, by a similar analysis than that we made in
type I and II models.

$\bullet$ Type I runaway behavior: If $N_M(N_R) > 2$ there is always
a runaway direction parameterized by $|X| \to \infty$, in which the
fields behave asymptotically as \[|\phi^{(r_k)}_{a_k}| \sim
|X|^{\frac{2 N_M(k) - N_M(N_R)}{2}}\ , \ \forall\ k, \ a_k  \ . \]

$\bullet$ Type II runaway behavior: If $N_L(N_R) > 1$ there is
always a runaway direction parameterized by $X\to 0$, in which the
fields behave asymptotically as \[ |\phi^{(r_k)}_{a_k}| \sim
|X|^{-\frac{2 N_L(k) - N_L(N_R) - 1}{2}}\ , \ \forall\  k,\ a_k \ .\]
Notice that this is the SUSY vacua limit, and in this case it is
related to a runaway direction.

If these requirements are not fulfilled there are no runaway
directions.

~

We have then discussed in this section rather general supersymmetric
models with chiral superfields in which R-symmetry is explicitly
broken. Our results can be summarized by stating that all the three
models exhibit runaway directions, and only type III models have
SUSY vacua. Moreover, we have argued that the non-supersymmetric
vacua can be longlived.

\section{Chiral superfields in $\bf 5 \oplus  \bf \bar5$ representation
of $SU(5)$}

Since we want to consider the coupling of matter to the Standard
Model non-Abelian gauge fields, we shall discuss here models with
chiral superfields in a non-singlet representation. In particular,
we shall consider the case of $N_\phi$ pairs of fields $\phi_i$,
$\ti \phi_i$ transforming in the $\bf 5 \oplus  \bf \bar5$
representation under $SU(5)$.  The gauge dynamics will not be turned
on, and the only difference with respect to the models considered in
the previous section is that now we have two independent set of
$N_\phi$ fields, with an additional $SU(5)$ index, which we will
omit in the notation.

The models are defined by a canonical K\"ahler potential and the
superpotential
\be W = f X + {\cal M}(X)_{ij}\ \phi_i\ti\phi_j\ \;\;\;\;\;\;\;\; i,j=1,2,\ldots, N_\phi\ .
\label{SuperpotEOGM} \ee
In order to compare with the results of the previous section
(O'Raifeartaigh like models
 with singlets)  we suppose that $\cal M$ is also of the form ${\cal M} = M + B + X(L  + A)$
 with $M,B,L,A$ analogous to those in  (\ref{SelectionRules})
\bea M_{ij} \neq 0 &\Rightarrow& R(\phi_i) +
R(\ti\phi_j) = 2\ , \nonumber \\
 L_{ij}\neq 0 &\Rightarrow& R(\phi_i) +
R(\ti\phi_j) = 2\ , \nonumber \\
B_{ij} \neq 0  &\Rightarrow& R(\phi_i) +
R(\ti\phi_j) < 2\ ,\nonumber\\ A_{ij} \neq 0  &\Rightarrow& R(\phi_i) +
R(\ti\phi_j) < 2\ .\label{RsymmBreaking}\eea
Note that in our case supersymmetry is not dynamically broken so
that all parameter scales are put in by hand. However, in
refs.\cite{Dine:2006xt},\cite{Dine:2006gm}-\cite{Dine:2007dz} it has been shown how to
retrofit O'Raifeartaigh models rendering their scales dynamically
(this is what should happen as realized in \cite{Witten:1981nf}).
One could then think to apply this procedure to our models in such a
way that the particular form of the mass-matrices
(\ref{RsymmBreaking}) is enforced by symmetries.

Recall that models defined by
(\ref{SuperpotEOGM})-(\ref{RsymmBreaking}) are not R-symmetric, for
the same reasons as in the singlet case in section 2. As in that
case, setting different subset of parameters to zero, one obtains
different R-symmetric models (and some reassignment of R-charges
might be necessary to check this).

The scalar potential resulting from (\ref{SuperpotEOGM}) is
\be V(\phi, \ti \phi) = | f + (L + A)_{ij}\phi_i \ti \phi_j|^2 + |
(M + B + X(L + A))_{ij}\phi_i |^2 + | (M + B + X(L + A))_{ij}\ti
\phi_j |^2\ ,\ee
and non-supersymmetric extrema take place  at
\be \phi_i = \ti \phi_i = 0 \ , \ \ \ \ \ F = f \, ,
\ \ \ \ \
\forall X \ ,
\label{esesta}
 \ee
where $V= |f|^2$. The mass matrices take the same form as in
(\ref{Masas})-(\ref{masas2}). The extrema correspond to stable minima when
the bosonic mass-matrix
\be {\cal M}_B^2 = \left(\begin{matrix} {\cal M}^\dag {\cal M} & \bar f (L + A)^\dag \\
f (L + A) & {\cal M} {\cal M}^\dag
\end{matrix}\right) ,\ee
has no tachyonic eigenvalues. In addition there can be other
non-SUSY vacua elsewhere is field space. Concerning supersymmetric
vacua, the F-term conditions read
\bea f + (L + A)_{ij}\ \phi_i \ti\phi_j &=& 0\ ,\label{EOGM1}\\
{\cal M}(X)_{ij} \ \phi_i &=& 0\ ,\label{EOGM2}\\
{\cal M}(X)_{ij} \ \ti\phi_j &=& 0\ .\label{EOGM3}\eea
These equations are solvable only in the limit in which $\det{\cal
M}(X) = 0$.

Before discussing in detail the landscape of SUSY vacua for the
different type of models, let us mention some general features.
First of all, let us stress   that for a fixed $f$,  taking $L$ and
$A$ sufficiently small the solution of eq. (\ref{EOGM1}) will lead
to values of  $\phi_i, \tilde \phi_j $ sufficiently far away from
the non-SUSY vacua  $\phi_i = \tilde \phi_j = 0$. As wee will see, in general one can take small entries for $L$ and $A$ without destabilizing the non-supersymmetric vacuum, and thus one can in principle make it longlived.

As already mentioned, the existence of SUSY vacua requires  non-zero
configurations for $\phi$, $\ti \phi$ in order to solve
(\ref{EOGM1}). If this is the case, one can start by solving
 equations (\ref{EOGM2}), (\ref{EOGM3}) separately.   Since $\det {\cal M} =
\det{\cal M}^T$, either each one corresponds to a non-trivial
solution or both lead to
 $\phi_i = \tilde \phi_i = 0$. Now, since both can be cast in the
form studied in the singlet case,  nontrivial solutions can be
inferred from those discussed in the previous section.

All the models we are  considering will  have $\det(M + X L) \neq 0$ for some
values of $X$. Then, as we explain in the Appendix (where we also define the notation), given a basis in
which the fields are ordered by increasing R-charge, $M + X L$ will be
anti-diagonal by blocks with non-zero determinant  and the fields  will
come in pairs with R-charges $r_i + \ti r_{N_R - i + 1} = 2$. In
this basis $\cal M$ has the form (see Appendix)
\be {\cal M}(X) = \left(\begin{matrix} (B + X A)^{(r_1, \ti r_1)} &
&
\dots & & (B + X A)^{(r_1, \ti r_{N_R-1})} & (M + X L)^{(r_1, \ti r_{N_R})} \\  &  & & & & 0 \\
 \vdots & & \dots & & & \\ & & & &\ \ \ \ \ \ \ \dots & \vdots\\
(B + X A)^{(r_{N_R-1}, \ti r_1)} & & & & & \\
(M + X L)^{(r_{N_R}, \ti r_1)} &  & 0& & \dots& 0\end{matrix}\right)\ .
\label{AntiDiagonal} \ee
It is clear from this equation that
\be
\det(B + X A) = 0\ , \ \ \ \ \ \ \det(M + B + X (L + A)) =
\det(M + X L)\ .
\ee

Let us classify the models as in the previous section, according to
the properties of matrix $\cal M$ , describing  SUSY vacua and
runaway behavior in each family.

~

\noindent{\bf Type I models}

The  non-supersymmetric configuration $\phi = \ti \phi = 0$ is a
stable minimum in some region $|X| < X_{max}$ when $f \ll
eigenvalues(\hat M ^2)$. SUSY is everywhere broken because the
following SUSY vacua equations cannot be satisfied
\bea f + A_{ij}\ \phi_i \ti\phi_j &=& 0\ ,\nn\\
(M + B + A X)_{ij} \ \phi_i &=& 0\ ,\nn\\
(M + B + A X)_{ij} \ \ti\phi_j &=& 0\ .\label{EOGM11}\eea
The largest subset of equations that can be solved for non-vanishing $(\phi, \ti\phi)$ is
\bea f + A_{ij}\ \phi_i \ti\phi_j &=& 0\ ,\nn\\
(M + B + A X)_{ij} \ \phi_i &=& 0\ , \ \ \ \ \ \ R(\ti\phi_j) < \ti r_{N_R}\ ,\nn\\
(M + B + A X)_{ij} \ \ti\phi_j &=& 0\ , \ \ \ \ \ \ R(\phi_i) < r_{N_R}\ ,\label{EOGM1b}\eea
and are not compatible with the two remaining equations
\be M^{(r_{N_R},\ti r_1)} \ti \phi^{(\ti r_1)} = 0 \ , \ \ \  \ \
{(\phi^T)}^{(r_1)} M^{(r_1, \ti r_{N_R})}  = 0\ ,\ \ee
which  force $\phi = \tilde \phi = 0$. As in the previous section
there is a runaway direction when $|X| \to \infty$, where
\bea |\ti \phi^{(\ti r_k)}_{\ti a_k}| &\sim& |X|^{k-1} |\ti \phi^{(\ti r_1)}_{\ti a_1}| \ , \ \ \ \ti \phi^{(\ti r_1)} \to 0 \ , \ \ \ \ \forall k , \ti r_k, \ti a_k\ ,\nn\\
|\phi^{(r_k)}_{a_k}| &\sim& |X|^{k-1} |\phi^{(r_1)}_{a_1}| \ , \ \ \
\phi^{(r_1)} \to 0\ , \ \ \ \ \forall k , r_k, a_k\ , \eea
and   $\phi^{(r_1)}$ and $\ti\phi^{(\ti r_1)}$ remain undetermined
in opposition to the singlet case
\be |\phi^{(r_1)}_{a_1} \ti\phi^{(\ti r_1)}_{\ti a_1}| \sim |X|^{2 -
N_R}\ ,\ \ \ \ \forall a_1, \ti a_1\ .\label{Undet1}\ee
So, when $N_R > 2$, there is a continuous set of directions for
which there is an asymptotic SUSY vacua (In (\ref{Undet1}), $SU(5)$
indices are contracted). We call this a {\it  runaway valley} (In
the $|X|\to\infty$ direction). When $N_R = 1, 2$ there   is no
runaway behavior.

~

\noindent{\bf Type II models}

The $\phi = \ti \phi = 0$ non supersymmetric minima is stable
at some region $X_{min} < |X|$. In these
type of models SUSY is also broken, even if $X = 0$ which implies
$\det {\cal M} = 0$. The biggest subset of equations that can be
solved for non-vanishing $(X, \phi, \ti\phi)$ is
\bea f + (A + L)_{ij}\ \phi_i \ti\phi_j &=& 0\ ,\nn\\
(B + X A + X L)_{ij} \ \phi_i &=& 0\ , \ \ \ \ \ \ R(\ti\phi_j) < \ti r_{N_R}\ ,\nn\\
(B + X A + X L)_{ij} \ \ti\phi_j &=& 0\ , \ \ \ \ \ \ R(\phi_i) < r_{N_R}\ ,\label{EOGM2b}\eea
and these equations are incompatible with the two remaining equations
\be X L^{(r_{N_R}, \ti r_1)} \ti \phi^{(\ti r_1)} = 0 \ , \ \ \  \ \ {(\phi^T)}^{(r_1)} X L^{(r_1, \ti r_{N_R})}  = 0 \ ,
\ee
this forcing  $X\phi = X\tilde \phi = 0$. As in the singlet case,
there is a runaway direction when $X \to 0$, where
\bea |\ti \phi^{(\ti r_k)}_{\ti a_k}| &\sim& |X|^{1-k} |\ti \phi^{(\ti r_1)}_{\ti a_1}| \ , \ \ \ X \ti \phi^{(\ti r_1)} \to 0\ , \ \ \ \ \forall k , \ti r_k, \ti a_k\ , \nn\\
|\phi^{(r_k)}_{a_k}| &\sim& |X|^{1-k} |\phi^{(r_1)}_{a_1}| \ , \ \ \
X \phi^{(r_1)} \to 0\ , \ \ \ \ \forall k , r_k, a_k\ , \eea
and $\phi^{(r_1)}$ and $\ti\phi^{(\ti r_1)}$ are again undetermined
\be |\phi^{(r_1)}_{a_1} \ti\phi^{(\ti r_1)}_{\ti a_1}| \sim |X|^{N_R
- 1}\ ,\ \ \ \forall a_1, \ti a_1\ .\ee
When $N_R > 1$ there is always a runaway valley in the $X\to 0$ direction,
otherwise there is not.

~

\noindent{\bf Type III models}

These models have a stable non-supersymmetric vacuum in some
region $X_{min} < |X| < X_{max}$ when $f \ll eigenvalues(\hat M ^2)$, and also
(the notation is explained in the Appendix)
\begin{itemize}
\item Non-generic SUSY vacua if $N_M(1) = 0$ and $N_M(N_R-1) = N_M(N_R)$, and generic SUSY vacua if not.
\item Type I ($|X|\to\infty$) runaway valleys when $N_M(N_R) > 2$.
\item Type II ($X\to 0$) runaway valleys when $N_L(N_R) > 1$.
\item No runaways when $N_M(N_R) \leq 2$ or $N_L(N_R) \leq 1$.
\end{itemize}

\subsection{An explicit example}\label{Explicit}

\begin{figure}\begin{center}
\includegraphics[width=10 cm]{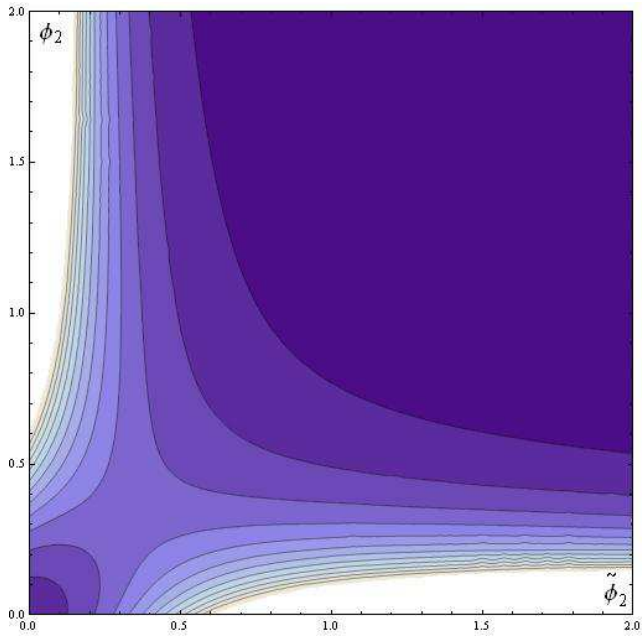}
\end{center}
\caption{\small A contour plot of the scalar potential for a type II
model of Section 3, taking all fields as real and $X ,  \phi_1$ and
$ \ti\phi_1$ in terms of $\phi_2$ and $\ti\phi_2$ according to
eq.(\ref{59}). Darker regions correspond to lower values of the
scalar potential. At the origin one can see a non-supersymmetric
vacum and far from the origin  a runaway valley.} \label{Valley}
\end{figure}

\begin{figure}\begin{center}
\includegraphics[width=12 cm]{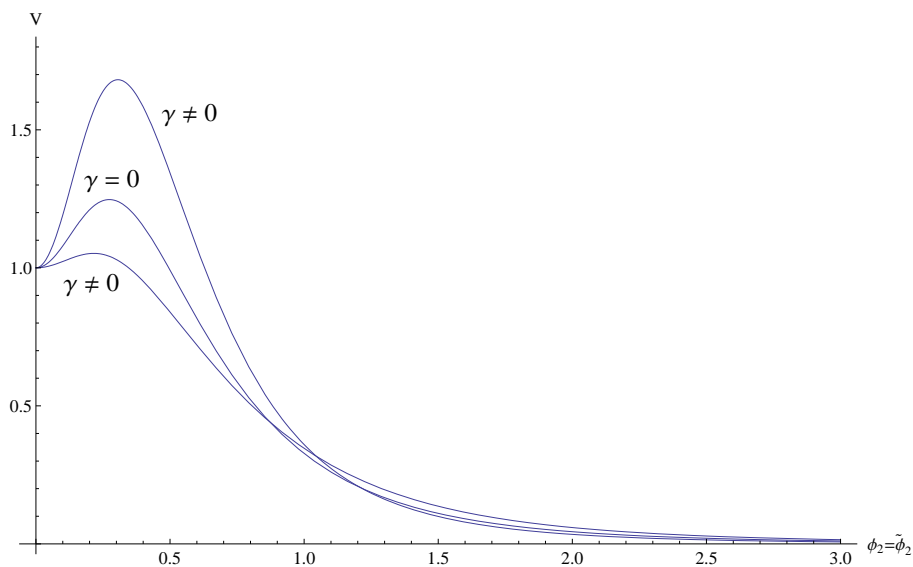}
\end{center}
\caption{A 2-dimensional plot of the type II scalar potential for
$\phi_2 = \tilde \phi_2$. The parameter $\gamma$ controls R-symmetry
breaking ($\gamma = 0$ corresponds to the R-symmetric case). We have
set $\eta = 1$, and guaranteed stability in all directions of
$\phi$-field space.} \label{perfiles}
\end{figure}

We shall now illustrate  the results above  concerning   runaway
directions and valleys by studying a specific example. We
consider a $N_\phi = 2$ type II model in the non-singlet case, defined
by the superpotential
\be W_{SU(5)} = f X + (a X  + b)\phi_1 \ti \phi_1 + \lambda X
(\phi_1 \ti \phi_2 + \ti \phi_1 \phi_2) = \eta x + (\gamma x + 1)
\phi_1 \ti \phi_1 + x (\phi_1 \ti \phi_2 + \ti \phi_1 \phi_2)\ .\ee
As before we have defined $x \equiv \lambda X / b\ , \ \gamma \equiv
a/\lambda$ and $\eta \equiv f /\lambda$; for definiteness we have
set $b = 1$. If one assigns the following R-charges $R(x) = 2\ ,\
R(\phi_1) = R(\ti \phi_1)= 1\ , \ R(\phi_2) = R(\ti \phi_2) = -1$
then parameter $\gamma$ will control R symmetry breaking, with
$\gamma = 0$ corresponding to the R-symmetric case.

The scalar
potential  reads (with the  $SU(5)$ indices    omitted),
\bea V_{SU(5)} &=& |1 + \gamma \phi_1 \ti \phi_1 + \ti \phi_1 \phi_2 +
\phi_1 \ti \phi_2|^2 + |(\gamma x + 1)\ti \phi_1 + x\ti \phi_2|^2 \nn\\ &&
\ + |(\gamma x + 1) \phi_1 + x \phi_2|^2 +  |x \phi_1|^2 + |x \ti
\phi_1|^2\ .\eea
Let us consider for definiteness a  fixed direction in field space,
such that only the first component in each multiplet is
non-vanishing
\be \phi^c_i = \phi_i (1, 0, \dots, 0)\ , \ \ \ \ti \phi^{\bar c}_i
= \ti \phi_i (1, 0, \dots, 0)\ ,\ee
where we have made explicit the   $SU(5)$ indices $c, \bar c$.  Along this direction the potential
slopes to zero through directions parameterized by $\ti \phi_2 \to
\infty$, $\phi_2 \to \infty$
\be (x, \phi_1, \ti \phi_1) \sim \left(\frac{1}{2\phi_2 \ti \phi_2},
-\frac{1}{2\ti \phi_2},-\frac{1}{2 \phi_2}\right)\ .\ee
Notice that  the direction in
field space defined by $\ti \phi_2 = h \phi_2$ is a runaway
direction for any value of a constant $h\neq 0$ and then one has  a continuous
set of runaway directions.   We sketch in figure \ref{Valley} the level
curves of this runaway valley, in the following direction in field space
\be (x, \phi_1, \ti \phi_1) = \left(\frac{5}{1 + 5 \phi_2 \ti
\phi_2}, -\frac{\phi_2}{1 + 2 \phi_2\ti \phi_2}, -\frac{\ti
\phi_2}{1 + 2 \phi_2\ti \phi_2}\right) \ .
\label{59}\ee
The metastable SUSY breaking vacua lies at the origin, where the boson
 mass-squared matrix reads
\be
m_B^2 = |b|^2 \left(\begin{matrix}|\gamma x + 1|^2 + |x|^2 & (\bar \gamma \bar x + 1) x & \bar y \bar \gamma  & \bar y  \\
(\gamma x + 1)\bar x & |x|^2 & \bar y  & 0 \\ y \gamma & y & |\gamma
x + 1|^2 + |x|^2 & ( \gamma x + 1) \bar x\\ y  & 0 & (\bar \gamma
\bar x + 1) x& |x|^2 \end{matrix}\right) \ , \ee
where we have defined $y \equiv \lambda f / |b|^2$. In the
limit of small $y$, for some $x > x_{min}$ the eigenvalues are all
positive in some range of $|\gamma|$. We plot in figure \ref{perfiles} the potential
$V_{SU(5)}$ along the curve $\phi_2 = \ti \phi_2$, which connects
the metastable vacua with a runaway direction for different values
of $\gamma$. For all these values, we have guaranteed stability in
all directions of field space. The figure shows how changing
$\gamma$ modifies the  well's depth and hence the lifetime of the
metastable vacua.  Our results suggest that with an appropriate
choice of the R-symmetry breaking parameter one  can increase the
lifetime of the metastable vacua (confront this with the case of
generic models, where explicit R-symmetry breaking generically
induces SUSY vacua lowering the lifetime of the metastable state).
Of course, in order to determine the non supersymmetric vacua's
life-time one should study quantum corrections, in the line of refs.
\cite{Riotto:1995am}-\cite{Dasgupta:1996pz}. Work on this issue is
in progress \cite{FMS}.

\section{Gauge mediating the supersymmetry breaking}

In this section we turn on the gauge dynamics so that the
non-singlet $\phi, \ti \phi$ messengers can interact through loops
with the MSSM fields. Concerning the singlet spurion field $X$, it
should acquire a non-vanishing F-component triggering SUSY breaking
\be X = X + \theta^2 F\ . \label{VEVX}\ee
The VEV of $X$ then gives mass to the messenger fields
through Yukawa-like superpotential terms
\be W_{mess} = {\cal M}(X)_{ij} \phi_i \ti \phi_j\ .
\label{MessSup}\ee
Supersymmetry breaking will then be communicated to the MSSM through
gauge interactions between $\phi, \ti \phi$ and the MSSM particles.
The scenario we have just described  corresponds to the mechanism
for gauge mediation of supersymmetry breaking (for details see for
example \cite{Giudice:1998bp}).

There are many possible ways in which the spurion $X$ develops the
VEV (\ref{VEVX}). In previous sections we considered the simplest
case, by adding a $fX$ term to the messenger superpotential. Of
course the addition of more complicated terms can also be
considered, but it will not be necessary for the analysis that
follows to specify one in particular, since we will only assume that
eq.(\ref{VEVX}) holds and focus on the messenger sector
(\ref{MessSup}).

Some extensions of minimal or ordinary gauge mediation (OGM) were
considered in \cite{Cheung:2007es}, with the messenger
superpotential containing all renormalizable couplings consistent
with the Standard Model gauge invariance, renormalizability, and
with a (spontaneously broken) R-symmetry, leading to a framework
which was called ``extra-ordinary gauge mediation'' (EOGM). In this
section we shall adopt the same strategy but in the case in which
R-symmetry is broken explicitly in the way discussed in section 3.
Following the route of \cite{Cheung:2007es}, we will show that the
explicit R-symmetry breaking terms included in the messenger sector
do not in general modify  the conclusions about the phenomenology of
EOGM. Minimal gauge mediation with superpotentials which are
deformed by mass terms were considered previously in
\cite{Dine:1981gu}-\cite{Dimopoulos:1982gm},
\cite{Izawa:1997gs}-\cite{Dimopoulos:1996yq}.

\subsection{Soft masses and effective messenger number}

An important property of some R-symmetric models is that the
determinant of the messenger mass matrix $\cal M$ is a monomial in
$X$ \cite{Cheung:2007es}. One can prove that this feature remains
valid in type I, II and III models with explicit breaking of
R-symmetry, where
\be \det {\cal M} = X^{n} G(M, L)\ , \ \ \ \ \ \ n =
\left\{\begin{matrix}
\!\!\!\!\!0\ \ \  \ \ \ \ \ \  \ \ \ \ \ \  \ \ \ \ \ \  \ \ \ \ \ \  \ \ \ \ \ \  \ \ \ \ \ \ \rm Type\ I \\ \!\!N_\phi  \ \ \ \ \ \ \ \ \ \ \ \ \  \ \ \ \ \ \  \ \ \ \ \ \  \ \ \ \ \ \ \ \ \ \ \ \ \rm Type\ II \\
\sum_{i=1}^{N_R} n_i (N_L(i) - N_L(i-1)) \ \ \ \ \ \rm Type\
III\end{matrix}\right.  \ .\label{DetIdent} \ee
We give the proof of this result in the Appendix.

The computation of gaugino and sfermion masses can be performed
generalizing the wavefunction renormalization technique
\cite{Giudice:1997ni}. Concerning gaugino masses, holomorphy allows
to substitute the VEV of the lowest component of $X$  by the $X$
superfield itself in the running coupling constant and renormalized
wave-function. In R-symmetric models, R-symmetry is
invoked to justify the analytical continuation $X \to \sqrt{X \bar
X}$ in the sfermion mass computation. We shall proceed in the same
way in the present case considering that the explicit R-symmetry
breaking represents just a small correction.

In this way, ignoring effects due to multiple messenger scales, one
finds for the gaugino and sfermion soft masses at the messenger
scale $M_{mess}$ to order ${\cal O} (F/M_{mess}^2)$
\cite{Cheung:2007es}
\bea
\tilde M_r &=& \frac{\alpha_r}{4\pi}\ \Lambda_G \ , \ \ \ \ \ \ \ \ \ \ \ \
\ \ \ \ \ \ \ \ \ \Lambda_G = F \partial_X \log {\det \cal M} = \frac{n F}{X}\ ,
 \label{GuaginoMasses}\\
m_{\tilde f}^2 &=& 2  \sum_{r=1}^3 C_{\tilde f}^r
\left(\frac{\alpha_r}{4\pi}\right)^2\ \Lambda_S^2 \ , \ \ \ \ \
\Lambda_S^2 = \frac{1}{2}F \bar F \frac{\partial^2}{\partial X
\partial \bar X}\sum_{i = 1}^N (\log |{\cal M}_i|^2)^2 \ . \label{SfermionMasses}\eea
Here ${\cal M}$ is the messenger mass matrix, ${\cal M}_i$ are
messenger masses, $C_{\tilde f}^r$ are the quadratic Casimir of
$\tilde f$ in the gauge group $r$, and ${\alpha_r}$ are the
messenger coupling constants.

In ordinary gauge mediation models the ratio
$\Lambda_G^2/\Lambda_S^2$ coincides with the number of messengers
$N_\phi$. This fact constraints the relation between gaugino and
sfermion masses, which is determined by the coupling constant. This
is no more valid in EOGM \cite{Cheung:2007es} nor in our
non-R-symmetric extension, where the ratio defines an $X$-dependent
effective messenger number $N_{eff}(X)$
\be N_{eff} ({\cal M}(X)) \equiv \frac{\Lambda_G^2}{\Lambda_S^2} \ ,
 \label{Neff}\ee
taking values between  $0 \leq N_{eff} \leq N_\phi$. For asymptotic
values $X \to 0$ and $X\to \infty$, the effective messenger number
$N_{eff}$ becomes independent of all the parameters in $A$, $B$, $L$
and $M$, and satisfies (see Appendix)
\bea
\frac{n^2}{n^2 - (N - r_m - 1)(2n - N + r_m)}\leq & N_{eff}(X\to 0)& \leq N - r_m \label{NeffRel1}\\
\frac{n^2}{r_\lambda + (r_\lambda - n)^2}\leq & N_{eff}(X\to
\infty)& \leq \frac{n^2}{r_\lambda + \frac{(r_\lambda - n)^2}{N -
r_\lambda}}\ , \label{NeffRel2}\eea
where we have defined
\be r_\lambda \equiv {\rm rank}\ (A + L) \ , \ \ \ \ r_m \equiv {\rm rank}\ (B + M)\ .\ee
In the most general case in which all possible  parameters
are indeed non-zero, the effective messenger number behaves
asymptotically as
\be N_{eff}(X \to 0) = \left\{\begin{matrix}
\!\!\!\!\!\!\!\!\!\!\!0 \ \ \ \ \ \ \rm Type\ I \\
1 \ \ \ \ \ \ \rm Type \ II, III
\end{matrix}\right.\ \ \ , \ \ \ \ \ \ \ \ \ N_{eff}(X \to \infty) = \left\{\begin{matrix}
\ \ \ \ \ \ \ \ 0 \ \ \ \ \ \ \ \ \ \ \ \ \ \ \ \rm Type\ I \\
\ \ \ \ \ \ \ \ N_\phi \ \ \ \ \ \ \ \ \ \ \ \ \ \ \rm Type \ II
\\ \frac{n^2}{N_\phi - 1 + (N_\phi - 1 - n)^2} \ \ \ \ \ \rm Type\ III\end{matrix}\right. \ .\label{LimNeff}\ee
One can infer from a specific example discussed in section 4.4 that
in the whole $X$-range and in the R-symmetric case, the $N_{eff}$
range for type II models is $1 - \delta \leq N_{eff} \leq N_\phi$
where $\delta \ll 1$, while for type III one has  $n^2 / (N_\phi - 1
+ (N_\phi - 1 - n)^2) <  N_{eff} \leq 1$. When R-symmetry is broken
these ranges can be extended. This is an interesting possibility
since the relation between gaugino and sfermion masses crucially
depend on $N_{eff}$. Moreover, when doublet/triplet splitting is
considered, sleptons could be taken to be lighter than squarks.
Finally, note that for type III models one can have $N_{eff} < 1$,
this opening the possibility of light gauginos (which takes place
for $N_{eff} \ll 1$), a fact that has interesting phenomenological
consequences \cite{Cheung:2007es}.

\subsection{Doublet/triplet splitting and unification}

Consider the superpotential with $SU(2)$ ($2 \oplus \bar 2$)
doublets $\ell, \ti \ell$ and $SU(3)$ ($3 \oplus \bar 3$) triplets
$q , \ti q$
\be W = {\cal M}^{(2)}_{ij}(X) \ell_i\ti \ell_j + {\cal
M}^{(3)}_{ij}(X) q_i\ti q_j\ .\ee
If we assume that doublets and triplets have the same R-charge
assignments, then equation (\ref{DetIdent}) holds for ${\cal
M}^{(2)}$ and ${\cal M}^{(3)}$ with the same number $n$. As a
consequence, equation (\ref{GuaginoMasses}) implies that the
relations between gaugino masses are preserved, independently of the
amount of doublet/tripplet splitting
\be M_1 : M_2 : M_3 = \alpha_1 : \alpha_2 : \alpha_3\ .
\label{gauginounifiaction}\ee

We then conclude that such  relations, that were believed to be
valid only for models with spontaneously broken $R$-symmetry, also
hold in these cases of explicit R-symmetry breaking\footnote{Breakdown of these relations is potentially problematic
concerning the electric dipole moment, due to the difference in the
phases of the gaugino masses.}. Since, as we shall see, unification of
the coupling constants at the GUT scale can be achieved in our
models, eq.(\ref{gauginounifiaction}) shows that   also gaugino mass
unification takes place.

The splitting we are considering also accounts for the sfermion masses
(\ref{SfermionMasses}) for which ${\Lambda_S^2}\to {\Lambda_S^2}_r$
with
\be {\Lambda_S^2}_r = \Lambda_G^2 {N_{eff}^{(r)}}^{-1} \ , \ \ \ \ r
= 2, 3\ . \label{smasssplit}\ee
Here ${\Lambda_S^2}_1 = \frac{2}{5}{\Lambda_S^2}_3 +
\frac{3}{5}{\Lambda_S^2}_2$, and we have defined ${N_{eff}^{(r)}}
\equiv N_{eff}({\cal M}^{(r)}(X))$. This shows that slepton and
squark masses (\ref{SfermionMasses}) are not only tied to the gauge
couplings $\alpha_r$, but also to the effective messenger number
${N_{eff}^{(r)}}$, thus leading to sfermion masses which are highly
modified with respect to the result of ordinary gauge mediation.
This effect can lead to small mass term $\mu$ and   explain  the
little hierarchy problem \cite{Cheung:2007es}. Equation
(\ref{smasssplit}) also receives contributions from the (non
positive definite) hypercharge D-terms $D_Y = g_Y (\phi^\dag Y_\phi
\phi - \ti \phi^T Y_\phi \bar \phi^*)$ which can drive slepton
masses to become tachyonic. To avoid this problem one can impose on
the model the   messenger parity
\cite{Dvali:1996cu}-\cite{Dimopoulos:1996ig}
\be \phi \to U^* \ti \phi^* \ , \ \ \ \ \ti \phi \to \ti U \phi^* \
, \ \ \ \ V \to - V\ ,\label{messparity1}\ee
where $V$ stands for the gauge superfields  and $U$ and $\ti U$ are
some unitary $N_\phi \times N_\phi$ matrices. This is a symmetry of
the Lagrangian provided the following conditions on the messenger
mass matrices hold
\be {\cal M}^\dag = U^\dag {\cal M} \ti U \ , \ \ \ \ \ ((L + A) F)^\dag
= U^\dag (L + A) F \ti U\ .\label{messparity2}\ee

Finally we notice that the splitting will also account for different
running of the coupling constants. Integrating the RG equations from
the ultraviolet scale $\Lambda$ down to the scale $\mu$ below the
lowest messenger scale gives
\be \alpha_r^{-1}(\mu) = \alpha_r^{-1}(\Lambda) + \frac{b_r -
N_\phi}{2\pi} \log\frac{\mu}{\Lambda} - \frac{1}{2\pi}\log \det
\frac{{\cal M}^{(r)}}{\mu}\ ,\ee
where $b_r = (-33/5, -1, 3)$ are the MSSM $\beta$-functions. If the
ultraviolet scale is the GUT scale $\Lambda = m_{GUT}$ one finds at
the electroweak scale $\mu = m_Z$ the following relation
\be \alpha_r^{-1}(m_{GUT}) = \alpha_r^{-1} (m_Z) + \frac{b_r}{2\pi}
\log\left(\frac{m_{GUT}}{m_Z}\right) -
\frac{N_\phi}{2\pi}\log\left(\frac{m_{GUT}}{\bar {\cal
M}^{(r)}}\right)\ ,\label{coupling}\ee
where we have defined $\bar {\cal M}^{(2, 3)} \equiv (\det{\cal
M}^{(2,3)})^{1/N\phi}$, and $\bar {\cal M}^{(1)} \equiv {\bar {\cal
M}^{(2)\ 3/5}} {\bar {\cal M}^{(3)\ 2/5}}$. When $\bar {\cal
M}^{(2)} = \bar {\cal M}^{(3)}$ unification is achieved at 1-loop as
in the MSSM, provided  the last term in (\ref{coupling}) vanishes
and the first two terms correspond precisely to the values of the
MSSM couplings at the GUT scale. The challenge is to achieve
arbitrary amount of doublet/triplet splitting ${\cal M}^{(2)} \neq
{\cal M}^{(3)}$ without spoiling unification (i.e., maintaining
$\bar {\cal M}^{(2)} = \bar {\cal M}^{(3)}$). In the R-symmetric
case, this is possible since the determinants of ${\cal M}^{(r)}$
are, in general, independent of some subset of the parameters of the
matrices ${\cal M}^{(r)}$. This subset can be bigger when R-symmetry
is explicitly broken, and moreover, the splitting can be produced
exclusively by R-symmetry breaking terms.

The relation $\bar {\cal M}^{(2)} = \bar {\cal M}^{(3)}$ implies
that both $SU(2)$ and $SU(3)$ sectors belong to the same type of
models (I, II, or III). In this case, the limits in which we found
runaway valleys or SUSY vacua (in the case in which the
superpotential was minimally completed with a $f X$ term, as in the
previous section) still hold for the complete model. Suppose on the
contrary that, for example, the $SU(2)$ sector is type I, and the
$SU(3)$ sector is type II so that $(\det M^{(2)})^{1/N_\phi} = \bar
{\cal M}^{(2)} \neq \bar {\cal M}^{(3)} = X (\det
L^{(3)})^{1/N_\phi}$. In this other case, there are neither SUSY
vacua nor runaway directions. The former is clear since type I and
II models have no SUSY vacua, while the latter holds because type I
and type II models have opposite ($X\to 0$ and $X\to \infty$
respectively) runaway directions and hence the potential slopes down
to zero in one sector while the other one tends to  a non zero
value.

\subsection{A comment on CP violating phases}

As it is well-known, sources of explicit CP violation can be
introduced in the MSSM through complex soft SUSY breaking terms. One
should then make sure that all the constants in the hidden sector
can be taken to be real by some re-phasing of fields, or otherwise
the phases of the couplings should be fine-tuned. Implementing this
condition together with that arising from messenger parity
(\ref{messparity1})-(\ref{messparity2}) in the R-symmetric case
highly restricts the form of the messenger sector (which has to be
necessarily fine-tuned), and such restriction is strengthen when
R-symmetry breaking terms are added since the messenger matrix $\cal
M$ can have far more entries to control.

Let us illustrate this difficulty with a simple example.
Consider type II models, first in the case in which parameters are chosen
so that R symmetry is not broken and the messenger superpotential reads
\be W_{mess} = \lambda X \sum_{i = 1}^{N_\phi} \phi_{N_\phi - i + 1} \ti \phi_i + m
\sum_{i = 1}^{N_\phi - 1} \phi_{N_\phi - i} \ti \phi_i \ .\ee
Let us define field and parameter  phases   as
\be \varphi_{\lambda X} \equiv - {\rm phase}(\lambda X)\ , \ \ \
\varphi_m \equiv - {\rm phase}(m)\ , \ \ \ \varphi_i \equiv {\rm
phase}(\phi_i)\ , \ \ \ \ti\varphi_i \equiv {\rm phase}(\ti\phi_i)
\ee

Re-phasing the fields and setting the parameter phases   to
zero leads  to a solvable linear system of $2 N_\phi - 1$
equations for $2 N_\phi$ unknowns
\[
 \left(\begin{matrix}1 & -1& & & & & & \\ & 1 & -1& & & & & \\ & & & & & &  &
 \\ & & & & & \dots & & \\ & &  & & & &  & \\& & & & & & 1 & - 1\end{matrix}\right)
  \left(\begin{matrix}\varphi_1\\ \varphi_2 \\ \\ \vdots \\ \\ \varphi_{N_\phi}\end{matrix}\right)
   = \left(\begin{matrix}\varphi_m - \varphi_{\lambda X}\\ \varphi_m - \varphi_{\lambda X}\\
   \\ \vdots \\ \\ \varphi_m - \varphi_{\lambda X} \end{matrix}\right)
   \]

\be
  \ti \varphi_i =  \varphi_{\lambda X} - \varphi_{N_\phi - i + 1}\
  \, , \;\;\;\;\;\; i = 1, \ldots, N_{\phi} \ .\label{systemphase}
\ee

Notice that this was only possible since there are only two
constants $\lambda$ and $m$, and this particular form of the superpotential is not enforced by symmetries. Now,
there are two different types of R-breaking terms that we can add to
the hidden sector of these models

\begin{itemize}
\item Terms of the form
\be \Delta W^{\not R}_{mess} = \gamma\ X\ \phi_{N_\phi - i_0} \ti
\phi_{i_0} \  , \ee
\end{itemize}
which can only be added if  $\varphi_m = -{\rm phase} (\gamma X)$,
so that one can make $\gamma$ real.

\begin{itemize}
\item Terms of the form ($\gamma$ may or may not be proportional to $X$)
\be \Delta W^{\not R}_{mess} = \gamma \phi_{i_0} \ti \phi_{j_0} \
, \ \ \ \ \ \ \ \ 1 \leq i_0 \leq N_\phi - 2\ , \ \ \ 1 \leq j_0 \leq N_\phi - i_0 - 1\ . \ee
\end{itemize}
In this case, defining $\varphi_\gamma = - {\rm phase} (\gamma)$
we have to add the following equation to the system (\ref{systemphase})
\be
\varphi_{i_0} - \varphi_{N_\phi - j_0 + 1} = \varphi_\gamma - \varphi_{\lambda X}\ ,
\label{systemphase2}
\ee
and in this case the complete system
(\ref{systemphase}),(\ref{systemphase2}) is   solvable only for the
following values of fine-tuned phases
\be
(N_\phi - j_0 - i_0 + 1) (\varphi_m - \varphi_{\lambda X}) = \varphi_\gamma - \varphi_{\lambda X}\ .
\ee

We then see that in the two cases one has to fine-tune  phases of
coupling constants  so that  they can be taken as real. Once reality
is imposed, eq.(\ref{messparity2}) can be satisfied (due to the fact that all $\lambda$'s and $m$'s are taken equal), and the matrices $U$ and $\ti U$ will depend on the added terms. In \cite{Carpenter:2008wi} the problem of obtaining messenger parity as an accidental symmetry is addressed.

\subsection{Explicit examples of type II and III models}

\begin{figure}\begin{center}
\includegraphics[width=14 cm]{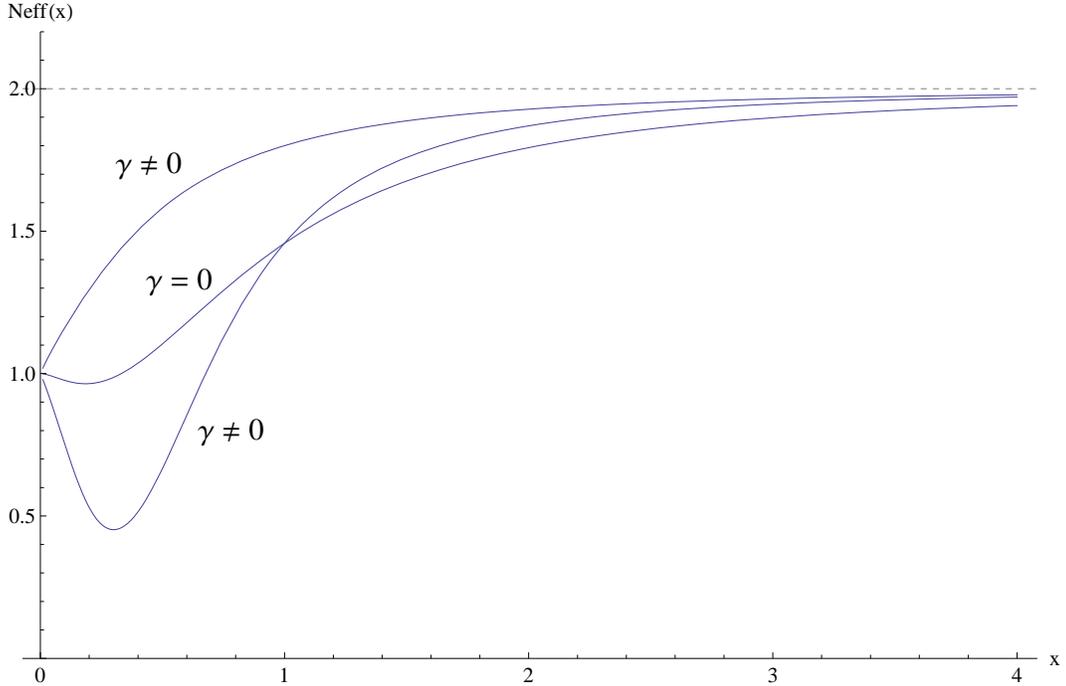}
\end{center}
\caption{A plot  of $N_{eff}(x)$ for different values of the
R-breaking parameter $\gamma$ in a type II model.} \label{NeffPlot2}
\end{figure}

As announced, we present here some explicit examples of gauge
mediation clarifying the previously discussed  features.  We will
not consider type I models since, as already pointed, lead to zero
gaugino masses to lower order in $F/M_{mess}^2$ (Almost all gauge
mediation models  in the literature in which the hidden sector is an
O'Raifeartaigh-like model  correspond to this type
\cite{Dine:1982zb}-\cite{Dimopoulos:1982gm},
\cite{Izawa:1997gs}-\cite{Nomura:1997uu},
\cite{Kitano:2006wm}-\cite{Xu:2007az}). Instead, we will explore
type II and III models where, as we have seen, gauginos are massive.
Let us start by considering the messenger sector of the type II example considered in
section \ref{Explicit}, with 2 messengers and superpotential
\be W_{mess} = (a X  + b)\phi_1 \ti \phi_1 +
 \lambda X (\phi_1 \ti \phi_2 + \ti \phi_1 \phi_2) = (\gamma x + 1)
\phi_1 \ti \phi_1 + x (\phi_1 \ti \phi_2 + \ti \phi_1 \phi_2)\ .
\ee
As before we have defined $x \equiv \lambda X / b\ , \ \gamma \equiv
a/\lambda$, and  set $b = 1$. Being R symmetry broken, R charge
assignment is arbitrary, and for definiteness  we chose
\[ R(x) = 2\ ,\ R(\phi_1) = R(\ti \phi_1)= 1\ , \ R(\phi_2)
= R(\ti \phi_2) = -1 \ .
\]
It is then clear that when $\gamma = 0$, the theory is R-symmetric.
Then, $\gamma$ will be taken as the parameter measuring the amount
of R symmetry breaking.

 As we have seen, at the classical level there are
non-supersymmetric vacua at $\phi = \ti \phi = 0\ ,\ \forall\ X$,
and a runaway valley at $X \to 0$. Computing the one loop quantum correction
to the moduli space
one should see that, as in the R-symmetric case, the $X$ field
acquires a VEV away from the origin ($X \ne 0 \Rightarrow x = (\lambda X/b)  \neq 0$),
at least for small $\gamma$ deformation.

The effective messenger number has two interesting limits
which are independent of  $\gamma$
\be \lim_{x \to 0} N_{eff}(x) = 1 \ , \ \ \ \ \lim_{x \to \infty} N_{eff}(x) = 2 \ .
\ee
However, the actual profile of $N_{eff} = N_{eff}(x)$ depends on
$\gamma$ as can be seen in figure \ref{NeffPlot2}. $N_{eff}$ can
take, in a certain $x$-range and for some values of $\gamma$, values
below $N_{eff} < 1$. Such enhancement in the range of values that
$N_{eff}(x)$ can take implies that the difference between the
effective numbers of the $SU(2)$ and $SU(3)$ sectors can be larger
than that resulting in the R-symmetric case. Indeed, here one can
have for certain values $x_2$ and $x_3$ that $|N_{eff}^{(2)}(x_2)-
N_{eff}^{(3)}(x_3)|>1$, in which case the slepton mass could be in
principle larger than the squark mass.

\begin{figure}\begin{center}
\includegraphics[width=14 cm]{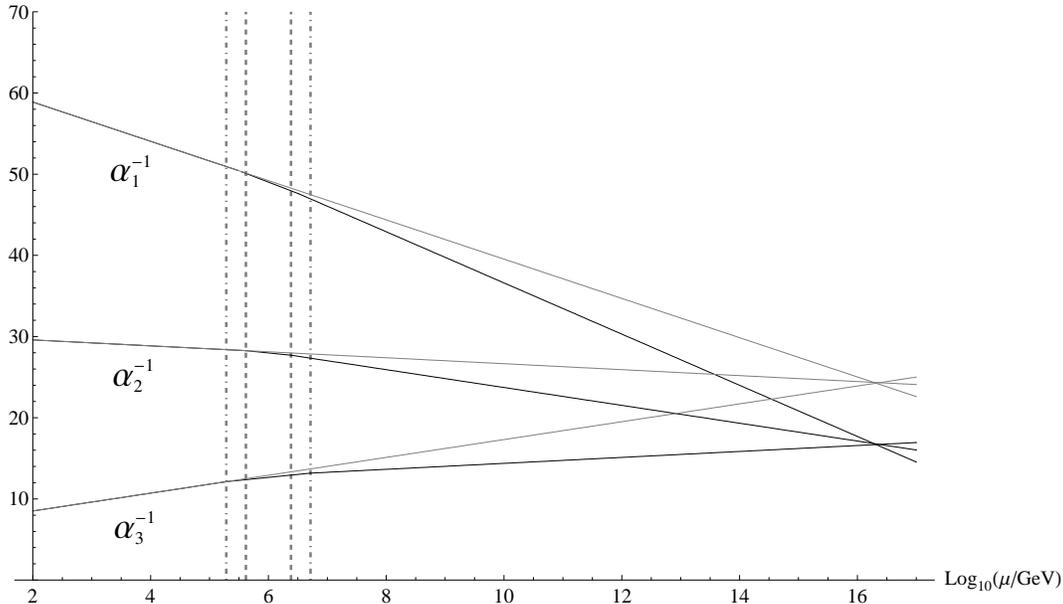}
\end{center}
\caption{The running of coupling constants is plotted both when the
hidden sector is present (darker full lines) and when it is absent
(lighter full lines).  Vertical lines indicate the mass scales of
the hidden sector. Dashed and  dot-dashed lines correspond to the
$SU(2)$ and $SU(3)$ sector in a two messenger model.}
\label{Unificacion}
\end{figure}

Now we turn to the study of the flow of the coupling constants,
considering doublet/triplet splitting in the messenger sector by
taking different mass matrices for $SU(2)$ and $SU(3)$. As explained
in section 4.2, R-symmetry can be broken differently in each sector
if $\gamma_2 \neq \gamma_3$ but
 unification can be achieved when
$\lambda_2 = \lambda_3$, independently of the values of $\gamma_i$.
We plot in figure \ref{Unificacion} the running of the couplings for
non-zero values of $\gamma_2$, $\gamma_3$, and take the masses of
the messengers between $200$Tev and $5200$Tev. We see that
unification takes place at the same scale as for the MSSM model with
a lower value of the unified coupling constant (the same as in
models with spontaneously broken R symmetry \cite{Cheung:2007es}).

\begin{figure}\begin{center}
\includegraphics[width=14 cm]{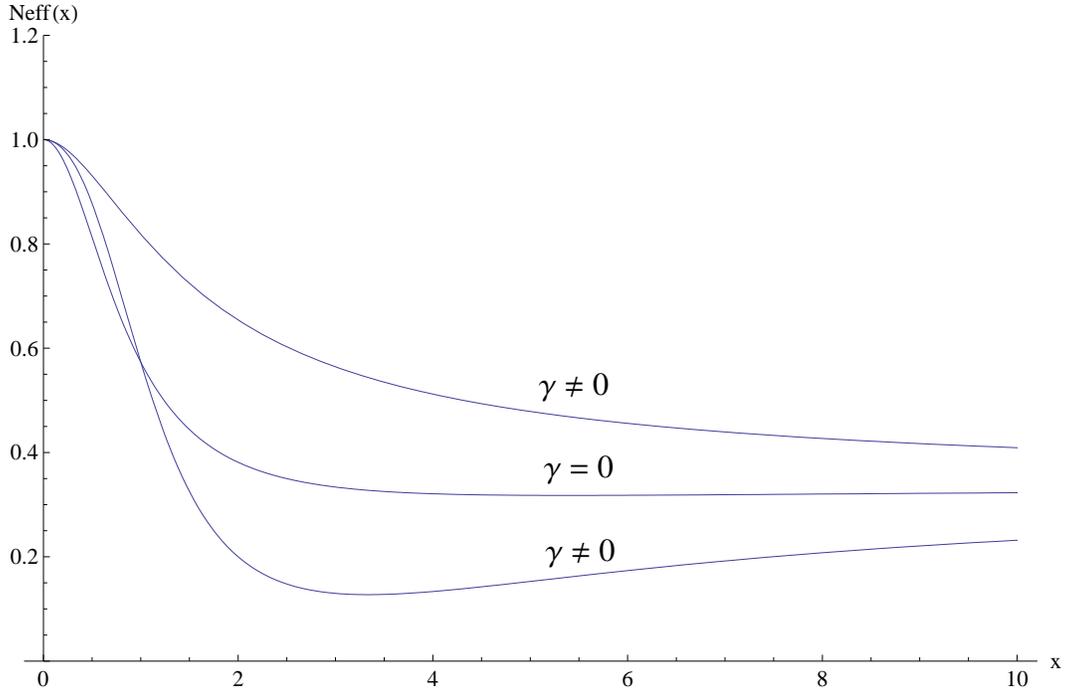}
\end{center}
\caption{A $N_{eff}(x)$ for different values of the R-symmetry
breaking parameter $\gamma$ in a type III model. Asymptotically all
three curves tend to $N_{eff}(\infty) = 2/5$.} \label{NeffPlot3}
\end{figure}

Let us end this section by discussing the simplest example of a type
III model, with $N_\phi = 3$ messengers, defined by the
superpotential
\be W_{mess} = \lambda' X \phi_1 \ti \phi_3 + m \phi_2 \ti \phi_2 +
m \phi_3 \ti \phi_1 + (a X + b) \phi_2 \ti \phi_1 = x \phi_1 \ti
\phi_3 + \phi_2 \ti \phi_2 + \phi_3 \ti \phi_1 + (x + \gamma) \phi_2
\ti \phi_1 \ .\ee
where we have defined $\gamma \equiv b/m$ and $x \equiv \lambda' X
/m$, and set $a = \lambda'$ and $m = 1$.

All calculations are analogous to those described in detail for the
type II case. As can be seen in figure \ref{NeffPlot3} the $\lim_{x
\to 0} N_{eff}(x) =1$, thus coinciding with the type II behavior. In
contrast, $\lim_{x \to \infty} N_{eff}(x) <1$, while in the type II
case $\lim_{x \to \infty} N_{eff}(x) > 1$. As in the type II model,
we achieve here a spread $|N_{eff}^{(2)}(x_2) - N_{eff}^{(3)}(x_3)|$
that  in the R-symmetric requires the addition of messengers. Note
that this spread is smaller than in the type II model previously
analyzed. Finally, unification of coupling constants is analogous to
that described for the type II model.

\section{Summary and discussion}

 We have analyzed, in the framework of gauge mediated supersymmetry
breaking, several families  of O'Raifear\-taigh-type models with
explicit R-symmetry breaking.

First we considered the simpler case of singlet chiral fields
 and determined the conditions under which such (non-generic) models  have SUSY vacua
and runaway behavior, in addition to a classical moduli of
non-supersymmetric minima which are therefore metastable.  Being the
models non-generic, one can have both longlived metastable vacua and
explicit, not necessarily small,  R-symmetry breaking. This is not
the case for generic models  where R-symmetry breaking implies the
existence of generic SUSY vacua and then one needs a small explicit
breaking in order to have longlived local minima.

 In order to
analyze phenomenological features of gauge mediation, we extended
the   analysis to the case in which the messenger fields are in
non-trivial representations of $SU(5)$ so that the resulting models
can be taken as candidates for the hidden sector that gauge mediate
SUSY breaking to the MSSM. The study of the corresponding vacua
landscape shows that some models (only type III) have SUSY vacua
only in one point of field space, except when the parameters are
fine-tuned, in which case there can be more than one supersymmetric
vacua. All models have continuously connected runaway directions,
the runaway valleys. In addition, there is a local SUSY breaking
minima at the origin of messenger field space which can be
longlived.

By gauging the   $SU(5)$ symmetry, we were able to analyze several
issues of gauge mediation, focusing on the impact of the explicit
breaking of R-symmetry. Considering a doublet/triplet-split of the
gauge group we found that the relations (\ref{gauginounifiaction})
between gaugino masses and coupling constants are satisfied. We then
showed that unification of couplings at the GUT scale is achieved in
the same way as in the MSSM, but for a different value of unified
couplings. As it happens in the R-symmetric case, in order to avoid
dangerous complex phases in the soft terms, we had to tune the
values of the coupling constants, which in addition allows to impose
a messenger parity. Finally, we presented arguments indicating that
slpetons can have masses lighter than those of squarks, as a result
of the R-symmetry breaking.

An interesting feature of the models we discussed is that the
R-symmetry breaking terms, chosen so that no additional SUSY-vacua
exists, are precisely those that preserve the phenomenological
features of extraordinary gauge mediation  \cite{Cheung:2007es}.
Another issue to be  pointed concerns the possibility of spontaneous
R-symmetry restoration. In fact, we have seen that the non-SUSY
moduli space contains  a point $X = 0$ in which R-symmetry is
classically not broken so that it would be possible
 that such point could correspond to a minimum also at the quantum level.

The breaking of supersymmetry that we discussed in our paper is not
dynamical. Following the proposal of
refs.\cite{Dine:2006gm}-\cite{Dine:2007dz}, it should be possible to
retrofit the models that we constructed so that  all mass scales
were generated dynamically. In this respect, the procedure should be
applied    in such a way that the particular form of the
mass-matrices (\ref{RsymmBreaking}) is imposed by the symmetries of
the dynamical SUSY breaking theory. Also, following the proposal
in ref.~\cite{Carpenter:2008wi}, it may be possible to achieve
messenger parity as an accidental symmetry. We hope to return to
these issues in a forthcoming work.

\section*{Appendix}
\subsection*{~A.1 Some identities for the
messenger mass matrices}\label{DetermCond}

Consider three (messenger mass) matrices $\Pi$, $\Omega_U$, and
$\Omega_D$ appearing in the superpotential in the form $(\Pi +
\Omega_U + \Omega_D)_{ij} \phi_i \ti \phi_j$, with $\det \Pi \neq 0$
and satisfying the following selection rules
\bea \Pi_{ij} &\neq& 0\ \Rightarrow\ R(\phi_i) + R(\ti \phi_j) = 2 \ ,\\
(\Omega_U)_{ij} &\neq& 0\ \Rightarrow\ R(\phi_i) + R(\ti \phi_j) < 2 \ ,\label{OmegaU}\\
(\Omega_D)_{ij} &\neq& 0\ \Rightarrow\ R(\phi_i) + R(\ti \phi_j) >
2\label{OmegaD} .\eea
We label fields so that $i<j \Rightarrow R(\phi_i)\leq R(\phi_j)\ ,
\ti R(\phi_i)\leq \ti R(\phi_j)$.

There are $(n_1, \dots, n_{N_R})$ fields with R-charge assignments
$(r_1, \dots, r_{N_R})$ and $(\ti n_1, \dots, \ti n_{\ti N_R})$
fields with R-charges $(\ti r_1, \dots, \ti r_{\ti N_R})$. We group
them in two sets   of $N_R$ and $\ti N_R$ vectors,
$\phi^{(r_i)}_{a_i}$ and $\ti \phi^{(\ti r_i)}_{\ti a_i}$,
respectively having $n_i$ and $\ti n_i$ components ($a_i =
1,\dots,n_i$; $\ti a_i = 1,\dots,\ti n_i$). Matrices $\Pi$,
$\Omega_U$, and $\Omega_D$ can then be arranged in blocks of $n_i$
rows and $\ti n_j$ columns each labeled by R-charges, and will be
denoted as $(\Pi + \Omega_U + \Omega_D)^{(r_i, \ti r_j)}_{a_i, \ti
a_j}$. We will omit indices $a_i, \ti a_j$ to simplify notation.

In this basis where fields are ordered by increasing R-charge, we
have
\be \Pi =
\left(\begin{matrix}0 & & & & \Pi^{(r_1, \ti r_{\ti N_R})} \\ & & & &  \\
& & \dots& & \\ & & & & \\  \Pi^{(r_{N_R}, \ti r_1)} & & & &
0\end{matrix}\right)\ . \label{FormaPi}\ee
Here each block must have non-zero determinant so that the condition
$\det \Pi \neq 0$ holds. This in turn implies that $\ti N_R = N_R$,
$\ti n_i = n_{N_R - i + 1}$, $r_i + \ti r_{N_R - i + 1} = 2$ and the
following identity
\be \det \Pi  = \prod_{i = 1}^{N_R}\det\Pi^{(r_i, \ti r_{N_R - i +
1})} \neq 0\ . \ee
Notice that the fields must come in pairs $(\phi_i,\ti \phi_{N_R - i
+ 1})$ satisfying $R(\phi_i) + R(\ti \phi_{N_\phi - i + 1}) = 2$.

We see from equations (\ref{OmegaU}) and (\ref{OmegaD})  that in
the adopted basis  $\Omega_U$ and $\Omega_D$  take the form
\be \Omega_U = \left(\begin{matrix} \Omega_U^{(r_1, \ti r_1)} &
\dots
& \Omega_U^{(r_1, \ti r_{N_R-1})} & 0 \\ \vdots & \dots & & \\
\Omega_U^{(r_{N_R-1}, \ti r_1)} & & & \\ 0 & & &
0\end{matrix}\right)\ , \ \  \Omega_D = \left(\begin{matrix} 0 &
& & 0\\ & & & \Omega_D^{(r_{2}, \ti r_{N_R})} \\ &  & \dots & \vdots
 \\ 0 & \Omega_D^{(r_{N_R}, \ti r_2)}
& \dots & \Omega_D^{(r_{N_R}, \ti r_{N_R})}  \end{matrix}\right)\ .
\label{FormaOmega} \ee
It is now clear from (\ref{FormaPi}),(\ref{FormaOmega}) that the following identities hold
\be \det (\Pi + \Omega_U) = \det(\Pi + \Omega_D) = \det \Pi\ , \ \ \
\ \det(\Omega_D) = \det(\Omega_U) = 0\ . \ee
The analysis also holds in  case that the two set of fields are
identical, as in Section 2.

~

Let us now define \be \Pi \equiv M + X L \ .\ee In general, $\det
\Pi$ would be a degree $N_\phi$ polynomial in the variable $X$,
being $N_\phi$ the dimension of the matrix $\Pi$. However, by
restricting the matrices $M$ and $L$ we can force $\det \Pi$ to be a
monomial \be \det \Pi = X^n G(M , L)\ , \ee where $n$ is some
integer between $0 \leq n \leq N_\phi$. We consider 3 cases
\begin{itemize}
\item $\det M \neq 0$, $L = 0$. In this case $\det \Pi = \det M$ is
independent of $X$, so $n = 0$.

\item $\det L \neq 0$, $M = 0$. In this case
\be \det \Pi = \prod_{i = 1}^{N_R} X^{n_i} \det{L^{(r_i, \ti r_{N_R
- i + 1})}} = X^{\sum_{i = 1}^{N_R} n_i} \prod_{i = 1}^{N_R} \det
L^{(r_i, \ti r_{N_R - i + 1})} = X^{N_\phi} \det L\ , \ee
so $n = N_\phi$.

\item $M^{(r_i, \ti r_{N_R
- i + 1})}\neq 0 \Longleftrightarrow L^{(r_i, \ti r_{N_R - i + 1})}
= 0$, and also  $\det(M + L) \neq 0$.

{ Notice that these two facts imply
\be \det L^{(r_i, \ti r_{N_R - i + 1})}
\neq  0    \Longleftrightarrow   \det M^{(r_i, \ti r_{N_R
- i + 1})} =  0  \label{Primero}\ee

Defining $N_L(k)$ the number of non-zero
blocks $L^{(r_i, r_{N_R-i+1})}$ of the matrix $L$ ($i = 1, \dots,
k$), we can arrive to the conclusion that
\bea
\det L^{(r_i, \ti r_{N_R - i + 1})}
\neq  0 \Longleftrightarrow N_L(i) - N_L(i-1) = 1 \nn\\
\det L^{(r_i, \ti r_{N_R - i + 1})}
=  0 \Longleftrightarrow N_L(i) - N_L(i-1) = 0 \label{Segundo}
\eea

In this case,
\be \det \Pi = \prod_{i = 1}^{N_R} \left[\det M^{(r_i, \ti r_{N_R - i + 1})} +  X^{n_i} \det L^{(r_i, \ti r_{N_R - i +
1})}\right]
\ee
Now we can use (\ref{Primero}) and (\ref{Segundo}) to write
\bea
\det \Pi &=&  X^n \prod_{i = 1}^{N_R} \left[\det M^{(r_i, \ti r_{N_R - i + 1})} +  \det L^{(r_i, \ti r_{N_R - i +
1})}\right]\nn\\
&=& X^n \prod_{i = 1}^{N_R} \det \left[ M^{(r_i, \ti r_{N_R - i + 1})} + L^{(r_i, \ti r_{N_R - i +
1})}\right]\nn\eea
Then
\bea\det \Pi&=& X^n \det (M + L)\ , \label{DimAnalysis}\eea
where we have defined
\be n = \sum_{i = 1}^{N_R} n_i (N_L(i) - N_L(i -1)) \ .\ee
Although $M$ has dimensional entries and $L$ is adimensional, since they are non-overlapping matrices the determinant $\det (M + L)$ has well defined mass dimensions. In units in which the superpotential $[W] = [{\rm mass}]^3$ and fields $[\phi] = [X] = [{\rm mass}]$, we have $\det(M + L) = [{\rm mass}]^{N_\phi - n}$ and then both sides in (\ref{DimAnalysis}) have dimension [mass]$^{N_\phi}$.

}

\end{itemize}

Finally, let us notice  that calling $\Omega_U \equiv B + A X$, and
defining
\be r_\lambda \equiv A + L \ , \ \ \ \ r_m \equiv B + M\ ,\ee
we can  follow \cite{Cheung:2007es}, to derive the relations
(\ref{NeffRel1})-(\ref{NeffRel2}) for the effective messenger number
$N_{eff}$ defined in (\ref{Neff}). In order to do that we must
determine the dependence of the messenger mass matrix eigenvalues
with respect to the field $X$. In what follows, we take $A$ and $L$
entries to be of order ${\cal O}(\lambda)$, and $B$ and $M$ of order
${\cal O}(m)$.

For large $X$, there are $r_\lambda$ messengers having ${\cal
O}(\lambda  X)$ masses while the remaining $N_{\phi} - r_\lambda$
messengers have a mass scaling with smaller powers of $X$
\be {\cal M}_i \sim X^{-n_i} \ , \ \ \ \ n_i \geq 0\ , \ \ \ \
\sum_{i = 1}^{N_\phi - r_\lambda} n_i = r_\lambda - n \
.\label{RelaCalMi1}\ee
At small $X$, $r_m$ messengers have masses of the order ${\cal
O}(m)$, and the remaining $N_{\phi} - r_m$ messengers have
\be {\cal M}_i \sim X^{n_i' + 1} \ , \ \ \ \ n_i' \geq 0\ , \ \ \ \
\sum_{i = 1}^{N_\phi - r_m} n_i'=  n - (N_\phi - r_m) \
.\label{RelaCalMi2}\ee
These identities imply $N - r_m \leq n \leq r_\lambda$ and
\be N_{eff}(X \to 0) = \frac{n^2}{\sum_{i = 1}^{N - r_m}(n_i' +
1)^2}  \ , \ \ \ \ \ \ N_{eff}(X \to \infty) = \frac{n^2}{r_\lambda
+ \sum_{i = 1}^{N-r_\lambda} n_i^2} \ .\label{AsyNeff}\ee
Equations (\ref{RelaCalMi1})-(\ref{AsyNeff}) can be combined to show
that the bounds (\ref{NeffRel1})-(\ref{NeffRel2}) hold.

\subsection*{A.2 ~Some simple examples of O'R models with singlets}

Here we analyze the classical landscape of the scalar potential in
some simple O'Raifeartaigh models, in order to make contact between
the general results of section 2 and explicit examples. Since we
have already analyzed type II models, we focus on type I and III.

~

\noindent {\bf Type I models}

The $N_\phi = N_R = 2$ case is the non-R-symmetric version of the
basic O'Raifeartaigh model \cite{O'Raifeartaigh:1975pr}
\be W_{(O'R)} = \frac{a}{2} X \phi_1^2 + \frac{b}{2} \phi_1^2 + m
\phi_1 \phi_2 + f X\ , \ee
where $b$ is an R-breaking parameter. This model has no SUSY vacua,
and since $N_R = 2$ there is no runaway behavior. Notice that a
shift $X \to X - b/a$ leads to the original O'Raifeartaigh model,
and then our results explain, in a general context, why this model
has no runaway directions.

The case $N_\phi = N_R = 3$ is the R-symmetry breaking version of a
R-symmetric model thoroughly analyzed in \cite{Shih:2007av}. We take
as  superpotential
\be W_{(Sh)} = \frac{a_1}{2} X \phi_1^2 + a_2 X \phi_1 \phi_2 +
\frac{b_1}{2} \phi_1^2 + b_2 \phi_1 \phi_2 + m_1 \phi_1 \phi_3 +
\frac{m_2}{2}\phi_2^2 + f X\ . \ee
We now show how fine-tuning different set of parameters one can
obtain different R-symmetric models.  Note that
\bea {\rm Setting}\ \ a_2 = b_1 = b_2 = 0\  & {\rm one\
must\ set} &  \ R(X) = R(\phi_3) = 2,\ R(\phi_1) = 0, \
R(\phi_2) = 1\nn\\
{\rm Setting}\ \ a_1 = b_1 = b_2 = 0\  & {\rm one\ must\
set} & \ R(X) = 2,\ R(\phi_1) = - R(\phi_2) = -1,\ R(\phi_3) =
3\ .\nn\ \eea
There is a runaway direction
\bea |X| \to \infty\ , \ \ \ \phi_1 \!\!\!&=&\!\!\!
\pm\sqrt{\frac{f}{\frac{a_2}{m_2}(a_2 X + b_2) - \frac{a_1}{2}}}\ ,
\ \ \ \phi_2 = \mp \frac{1}{m_2}\sqrt{\frac{f (a_2 X +
b_2)^2}{\frac{a_2}{m_2}(a_2 X + b_2) - \frac{a_1}{2}}}\ , \nonumber\\
\phi_3 \!\!\!&=&\!\!\! \mp \frac{1}{m_1}\sqrt{\frac{f
}{\frac{a_2}{m_2}(a_2 X + b_2) - \frac{a_1}{2}}}\left[a_1 X + b_1 -
\frac{1}{m_2} (a_2 X + b_2)^2\right]\ , \eea
which is asymptotically the same as that of the original
model\cite{Shih:2007av}. Along this runaway direction, the scalar
potential reaches the value of the SUSY breaking vacua ($V = |f|^2$)
at
\be X = \frac{2 m_1^2 m_2 + f a_1 m_2 - 2 f a_2 b_2}{2 f a_2^2} =
X_R + \frac{a_1 m_2 - 2 a_2 b_2}{2 a_2^2}  \ .\ee
Considering $a_2$ as a small parameter, we can make the non SUSY
minimum $\phi = 0$ stable near $X \sim 0$ for some range of
parameters and also sufficiently long lived.

~

\noindent{\bf Type III models}

Let us consider two $N_\phi = N_R = 3$ examples. One corresponding
to the superpotential
\be W = \frac{a_1}{2} X \phi_1^2 + a_2 X \phi_1 \phi_2 +
\frac{b_1}{2} \phi_1^2 + b_2 \phi_1 \phi_2 + m \phi_1 \phi_3 +
\frac{\lambda}{2} X \phi_2^2 + f X\ . \ee
Here, $N_L(N_R) = 1$ and $N_M(N_R) = 2$ and then  there are no
runaway directions. There is a SUSY vacua at
\be X = 0\ , \ \ \ \phi_1 = 0\ , \ \ \ \phi_2 = \mp i \sqrt{\frac{2
f}{\lambda}}\ , \ \ \ \phi_3 = \mp i \sqrt{\frac{2 f b_2^2}{\lambda
m^2}} \ ,\ee
which can be taken far from the non-SUSY $\phi = 0$ minima by
considering small $|\lambda|$.

The second example corresponds to   a superpotential of the form
\be W = \frac{a_1}{2} X \phi_1^2 + a_2 X \phi_1 \phi_2 +
\frac{b_1}{2} \phi_1^2 + b_2 \phi_1 \phi_2 + \frac{m}{2} \phi_2^2 +
\lambda X \phi_1 \phi_3 + f X\ , \ee
and $N_L(N_R) = 2$ and $N_M(N_R) = 1$. Then,   it has a runaway
behavior as that in Type II models,
\bea X \to 0\ , \ \ \ \phi_1 &=& \pm \sqrt{\frac{f X}{b_1 -
\frac{b_2^2}{m} + \left(\frac{a_1}{2} - \frac{a_2 b_2}{m}\right) X}
}\ , \ \ \ \phi_2 = \mp \frac{1}{m}\sqrt{\frac{f X (a_2 X +
b_2)^2}{b_1 - \frac{b_2^2}{m} + \left(\frac{a_1}{2} - \frac{a_2
b_2}{m}\right) X}
}\ , \nonumber \\
\phi_3 &=& \mp \frac{1}{\lambda X} \sqrt{\frac{f X}{b_1 -
\frac{b_2^2}{m} + \left(\frac{a_1}{2} - \frac{a_2 b_2}{m}\right) X}
}\left[a_1 X + b_1 - \frac{1}{m}(a_2 X + b_2)^2\right]\ .\eea
In addition, when $X = 0$, for the fine-tuned values of the
couplings $b_1 m = b_2^2$, there is a one dimensional space of SUSY
vacua parameterized by $\phi^1$
\be \phi_2 = - \frac{b_2}{m} \phi_1 \ , \ \ \ \phi_3 = -
\frac{1}{\lambda \phi_1} \left(f + \left(\frac{a_1}{2} -
\frac{b_2}{m}\right) \phi_1^2\right) \ .\ee

\vspace{1 cm} \noindent\underline{Acknowledgements}: This work was
partially supported by UNLP, UBA, CICBA, CONICET and ANPCYT.
\newpage

\end{document}